\documentclass[aps,amsfonts,reprint,tightenlines,amssymb,superscriptaddress,twocolumn]{revtex4-1}
\usepackage[T1]{fontenc}    
\usepackage{amsfonts} 
\usepackage{amsmath,amsbsy,amssymb,graphicx,enumerate,latexsym,bm,ulem,multirow,listings}                       
\usepackage{times} 
\usepackage{color}

\usepackage[pagebackref=false]{hyperref} 
\hypersetup{
    bookmarks=true,         
    unicode=false,          
    pdftoolbar=true,        
    pdfmenubar=true,        
    pdffitwindow=false,     
    pdfstartview={FitH},    
    pdfkeywords={keyword1} {key2} {key3}, 
    pdfnewwindow=true,      
    colorlinks=true,       
    linkcolor=red,          
    citecolor=blue,        
    filecolor=magenta,      
    urlcolor=cyan,           
}

\makeatletter
\newsavebox{\@brx}
\newcommand{\llangle}[1][]{\savebox{\@brx}{\(\m@th{#1\langle}\)}%
  \mathopen{\copy\@brx\kern-0.5\wd\@brx\usebox{\@brx}}}
\newcommand{\rrangle}[1][]{\savebox{\@brx}{\(\m@th{#1\rangle}\)}%
  \mathclose{\copy\@brx\kern-0.5\wd\@brx\usebox{\@brx}}}
\makeatother

\begin{document}
\title{Negative-Temperature State Relaxation and Reservoir-Assisted Quantum Entanglement in Double Spin Domain Systems}               
\author{Yusuke~Hama} 
\affiliation{National Institute of Informatics, 2-1-2 Hitotsubashi, Chiyoda-ku, Tokyo 101-8430, Japan}
\author{Emi~Yukawa} 
\affiliation{RIKEN Center for Emergent Matter and Science (CEMS),Wako, Saitama, 351-0198, Japan}
\author{William~J.~Munro}  
\affiliation{NTT Basic Research Laboratories, NTT Corporation, 3-1 Morinosato-Wakamiya, Atsugi-shi, Kanagawa, 243-0198, Japan }        
\affiliation{National Institute of Informatics, 2-1-2 Hitotsubashi, Chiyoda-ku, Tokyo 101-8430, Japan}
\author{Kae~Nemoto}
\affiliation{National Institute of Informatics, 2-1-2 Hitotsubashi, Chiyoda-ku, Tokyo 101-8430, Japan}
\date{\today} 
\begin{abstract}{ 
Spin collective phenomena including superradiance are even today being intensively investigated with experimental tests performed based on state-of-the-art quantum technologies. Such attempts are not only for the simple experimental verification of predictions from the last century but also as a motivation to explore new applications of spin collective phenomena and the coherent control of the coupling between spin ensembles and reservoirs. In this paper, we investigate the open quantum dynamics of two spin ensembles (double spin domains) coupled to a common bosonic reservoir. We analyze in detail the dynamics of our collective state and its structure by focusing on both the symmetry and asymmetry of this coupled spin system.  We find that when the spin size of one of the double domains is larger than that of the other domain, at the steady state this system exhibits two novel collective behaviors: the negative-temperature state relaxation in the smaller spin domain and the reservoir-assisted quantum entanglement between the two domains. These results are the consequence of the asymmetry of this system and the decoherence driven by the common reservoir. }
\end{abstract}
\maketitle
\section{Introduction}\label{intro}
Our recent advances in material device fabrication as well as highly effective signal detection have allowed us to reach the stage where various gedanken experiments from the earlier stages of the quantum physics can be realized in the laboratory (these include, for instance, quantum interference using a double slit, Bose-Einstein condensation, and superradiance \cite{AZeilingerRMP,BECRMP, superradiance1}). We are now entering at the era where we can integrate multiple sub quantum systems together into a single multi-functional quantum system (hybrid quantum systems, for instance, atoms coupled to optical cavities and nitrogen-vacancy (NV) centers in diamond coupled to flux qubit in superconducting circuits) \cite{hybrid1,circuitqedreview1,hybrid2}. The engineering of the hybrid quantum systems have been performed in quite diverse systems using elements coming from condensed matter to atomic, molecular and optical systems \cite{hybrid1,circuitqedreview1,hybrid2,putz2016,strongcoupling1,strongcoupling2,trappedionsreview1,trappedionsreview1,cavityqedreview1,cavityqedreview2,quantumdotreview1,nvcenterreview1,nvcentermechresonatorreview1,circuitqedreview1,nvcentermechresonatorreview1,mechanicalresonatorreview1}. Such multi-functionality of these hybrid quantum systems is superior to the functionalities of any individual systems \cite{hybrid1,circuitqedreview1,hybrid2,putz2016,qubus}. These developments have paved the way to allows us to explore novel phenomena in many-body and non-equilibrium quantum physics inherent from the hybridization process. Further they may allow new  and novel techniques for performing the quantum information processing.

One of the major focusses in hybrid quantum physics is the exploration of collective phenomena motivated by spin ensembles being coherently or collectively coupled to bosonic modes \cite{hybrid1,circuitqedreview1,hybrid2,putz2016,cstrongcouplingnatphys2014,cstrongcouplingCRphysique2016}. When a spin ensemble couples collectively to bosons, it shows stronger coupling than that between individual spin and bosons, which scales with the square root of the total spin number \cite{hybrid1,circuitqedreview1,hybrid2,putz2016,cstrongcouplingnatphys2014,cstrongcouplingCRphysique2016}.  The dynamics are characterized by this spin number $N$ (the size of spin ensemble) and are generally very different from the single-spin dynamics. The typical example is the superradiance where the spin ensemble shows extremely rapid decay on a timescale of $1/N$ with the strong radiative intensity also scaling with $N^2$ \cite{superradiance1,GH82,solidsuperradiance2016}.  Although it was proposed by Dicke over 60 years ago \cite{superradiance1}, superradiance and such collective quantum phenomena remain as both fascinating and important research fields in various systems using the state-of-the-art quantum technologies such as cavity quantum electrodynamical systems with atomic, molecular and optical setups \cite{cavityqedreview2} and solids  \cite{solidsuperradiance2016}.

Most prior research in superradiance has, however, focused on this collective phenomena with a single spin ensemble. We are now able to design and fabricate devices with multiple ensembles present on them. The next step is to analyze collective phenomena generated by the multiple spin ensembles and explore ways to control the coupling structure between multiple spin ensembles and the reservoirs. Such investigations will be important and interesting from two reasons. First and foremost since the dynamics of a single spin and those of the collective spin are radically different as in the case of the superradiance, we expect the nontrivial dynamics of multiple spin ensembles to arise owing to its complicated structure.  Second, collective spins form a strong coupling between bosons, which is going to be an important ingredient for quantum information processing \cite{hybrid1,circuitqedreview1,hybrid2,putz2016,cstrongcouplingnatphys2014,cstrongcouplingCRphysique2016}. The novel spin collective phenomena are staring to emerge in various experimental setups coherently controlling multiple spin ensembles and the reservoir \cite{DNP1,DNP2,Kumadaetal,Fauziprb,yhamaetal,thomas}. Towards these goals, we investigate in this paper the dynamics of the system with two spin ensembles (double spin domains, for instance,
double nuclear spin domains in GaAs semiconductor \cite{DNP1,DNP2} and electron spin ensembles in NV centers in diamond \cite{thomas}) coupled to a common bosonic reservoir. We begin by examining what kind of collective phenomena and its associated steady state are induced by the common bosonic reservoir characterizing them by the two spin-ensemble (domain) sizes (the numbers of spins present in each of the domains). When the first spin-domain size is much larger than the second, the double spin domains relax to steady states exhibiting two novel features: First is that the small spin domain relaxes to the negative-temperature state where the average excited-state population is greater than 50\% \cite{negativeTprl}.  Second is the creation of quantum entanglement between the two domains (even though they are not directly coupled together). These phenomena are realized due to the asymmetry of the double spin domains and decoherence driven by the common reservoir. 

This paper is organized as follows. It begins in Sec. \ref{model} with our mathematical model of the double spin domains coupled to the common bosonic reservoir. Then in Sec. \ref{DMS} (which presents the main results of this paper) we discuss how to analyze the dynamics of our double spin domain system and its structure using a symmetry argument. In particular, we will investigate the steady state characterized by the sizes of two spin ensembles. We present two novel collective phenomena intrinsic to this system: the negative-temperature state relaxation of the smaller domain and reservoir-assisted quantum entanglement generated between the spin domains. In Sec. \ref{generalization}, we will present a generalization of the previous argument for larger spin systems. Finally in Sec. \ref{discussionconclusions}  we give a concluding discussion of this paper.

\section{Modeling}\label{model}
\begin{figure}[t] 
\includegraphics[width=0.4 \textwidth]{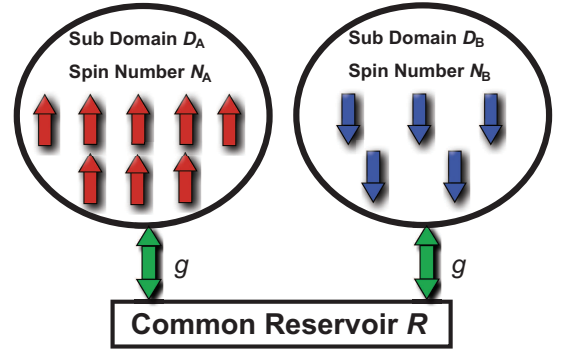}
\caption{The illustration of double spin domain system consists of two one-half spin ensembles and a common bosonic reservoir. 
The two spin domains couple equivalently to the common reservoir with a constant $g$ represented by two green arrows.  
The first domain $D_\text{A}$ has $N_\text{A}$ spins indicated by up red arrows, whereas the second domain $D_\text{B}$ contains $N_\text{B}$ spins described by blue arrows.
All the spins are identical.  } 
\label{doubledomain} 
\end{figure}

In this section, we present a mathematical model of our double spin domain system. 
As shown in Fig. \ref{doubledomain} it is a hybrid quantum system consists of two spin ensembles coupling to a common bosonic reservoir $R$ each with a coupling constant $g$. The temperature of the reservoir is $T$. 
Now let us name the first (second) domain as $D_{\text{A(B)}}$.  The domain $D_{\text{A(B)}}$ includes $N_{\text{A(B)}}$ individual spin 1/2 particles. All the spins in the double domain are identical species.   
The spin frequency is given by $\omega_{\text{s}}$.
Due to these conditions, both spin ensembles in the domains $D_{\text{A}}$ and $D_{\text{B}}$ couple to the common reservoir $R$ collectively and  these two spin ensembles act as collective spins 
$J^\alpha_{\text{A}}=\sum_{i_{\text{A}}=1}^{N_{\text{A}}} S^\alpha_{i_{\text{A}}}$ and $J^\alpha_{\text{B}}=\sum_{i_{\text{B}}=N_{\text{A}}+1}^{N_{\text{A}}+N_{\text{B}}} S^\alpha_{i_{\text{B}}}.$ 
Here $J^{\alpha}_a$ ($\alpha=x,y,z.$) are the collective spin operators for $x,y,z$ components of the domain $a$ ($a$=A,B) whose spin sizes are $N_{\text{A}}/2$ and $N_{\text{B}}/2$, respectively.  
$S^\alpha_{i_{\text{A}}}$ ($S^\alpha_{i_{\text{B}}}$) is the $i_{\text{A}}$-th ($i_{\text{B}}$-th) 1/2 spin operator.
Our combined system is described by the Hamiltonian
\begin{align}
H&  =
\hbar\omega_{\text{s}} (J^z_{\text{A}}+J^z_{\text{B}})+\int d^d k \;E_{\boldsymbol{k}}r^\dagger_{\boldsymbol{k}}r_{\boldsymbol{k}}\notag\\
&+\frac{\hbar g}{2} \left[(J^+_{\text{A}}+J^+_{\text{B}})R+(J^-_{\text{A}}+J^-_{\text{B}})R^\dagger\right].
\label{hamiltonian1} 
\end{align}
The first and second terms represent the Hamiltonian of the two spin domains and the common reservior $R$, respectively. 
The spin operators $J^{\pm}_a=J^x_a\pm iJ^y_a$ are the rising and lowering operators of  domain $a$. 
  $E_{\boldsymbol{k}}$ is the dispersion relation with $\boldsymbol{k}$,  its wavevector. We will take $E_{\boldsymbol{k}}$ to be linear. 
The dimension $d$ is the spatial dimension of this system while
$r _{\boldsymbol{k}}$ and $r^\dagger_{\boldsymbol{k}}$ are annihilation and creation operators of the reservoir, respectively.
They satisfy the commutation relation $[r _{\boldsymbol{k}},r^\dagger_{\boldsymbol{k}^\prime}]=\delta(\boldsymbol{k}-\boldsymbol{k}^\prime)$.
 The third term represents the interaction between the two spin domains and the common reservoir. 
  $R=\int  d^d k \kappa_{\boldsymbol{k}} r_{\boldsymbol{k}}$
  is the reservoir operator described by the annihilation operator $r_{\boldsymbol{k}}$ with a continuous function $\kappa_{\boldsymbol{k}}$. 
  The formula of $\kappa_{\boldsymbol{k}} $ is determined by the system we are considering.
  
 The dynamics we will analyze is the relaxation processes of the double spin domain induced by the reservoir $R$. 
 Such processes are described by the Lindblad master equation in the interaction picture as \cite{Carmichaeltxb} 
\begin{align}
\dot \rho(t)&=
\gamma\left[(\bar{n}+1){\cal L}(J^-_{\text{A}}+J^-_{\text{B}})+ \bar{n}{\cal L}(J^+_{\text{A}}+J^+_{\text{B}})\right]\rho(t),
\label{smasterequation1}
\end{align}
where the dot ``$\cdot$'' represents the time derivative and the Born-Markov approximation has been applied.  
The reduced density matrix $\rho$ is defined by tracing out the reservoir degrees of freedom over total density matrix as
$\rho(t)=$Tr$_{\text{R}}(\rho_{\text{tot}}(t))$. The reservoir density matrix is given by $\rho_{\text{R}}=\exp(- H_{\text{R}}/k_{\text{B}}T)$/Tr$_{\text{R}}$( $\exp(- H_{\text{R}} /k_{\text{B}}T)$) where $H_{\text{R}}$ is the second term in the total Hamiltonian   \eqref{hamiltonian1}  with $k_{\text{B}}$ the Boltzmann constant. The superoperator ${\cal L}(X)$ is defined by ${\cal L}(X)= 2X \rho X^\dagger - X^\dagger X \rho - \rho X^\dagger X$,  whereas 
$\gamma$ is the damping rate described by  the coupling $g$ and $|\kappa_{\boldsymbol{k}}|^2$ at the wavevector  $\boldsymbol{k}_{\text{s}}$ which satisfies $E_{\boldsymbol{k}_{\text{s}}}=\hbar\omega_{\text{s}}$. 
 $\bar{n}=1/(e^{\hbar\omega_{\text{s}} /k_{\text{b}}T}-1)$ is the Bose-Einstein distribution for the bosonic reservoir at the energy $\hbar\omega_{\text{s}}$. 
The first term in Eq. \eqref{smasterequation1} describes the absorption process of the spin ensembles while the second term represents the emission process.
In the following, we will demonstrate the relaxation processes at $T=0$.  For an initial state, we examine the anti-parallel configuration 
\begin{align}
\big{|}\text{is}\big{\rangle} 
&=|\uparrow\ldots\uparrow\rangle_{\text{A}}\otimes|\downarrow\ldots\downarrow\rangle_{\text{B}}.
\label{initialstate}
\end{align}
Here we choose the up (down)-spin state to be the excited (ground) state. The spin numbers are chosen such that $N_{\text{A}}\geq N_{\text{B}}.$
The relaxation processes in the double spin domain system are mathematically described by two expectation values 
$\langle  J_{\text{A}} \rangle=$Tr($\rho J_{\text{A}} $) and $\langle  J_{\text{B}} \rangle=$Tr($\rho J_{\text{B}} $). 
As we will see, the relaxation processes are the collective phenomena described by the two spin sizes $N_{\text{A}}$ and $N_{\text{B}}.$

Experimentally, the double spin domain system presented in Fig. \ref{doubledomain} can be realized in, for instance, QH system as a GaAs semiconductor.
In this system, nuclear spins can be initially polarized via the dynamic nuclear polarization (DNP) and form double spin domains consists of up-spin domain and down-spin domain \cite{DNP1,DNP2}.
On the other hand, the Nambu-Goldstone (NG) mode as a collective excitation of electron spin exhibiting a linear dispersion relation can be driven,
and the nuclear spins can couple to this NG mode which may play a role of the reservoir \cite{Kumadaetal,Fauziprb,yhamaetal}. 
By combining these two setups, we can prepare the double spin domain system shown in Fig. \ref{doubledomain}.
Another example is two electron spin ensembles in nitrogen-vacancy (NV) center in diamonds coupling to a superconducting resonator \cite{thomas}.

\section{System Structure}\label{DMS}
In this section, we will present the dynamics and the structure of the reduced density matrix $\rho$ for the double spin domain \eqref{initialstate} via the master equation \eqref{smasterequation1}.
In particular, we will analyze in detail the structure of the steady state characterized by the two spin sizes. 
For a preparation, we will first introduce a tensor-product spin space, a direct-sum spin space, and explain their relations.
We will solve the master equation \eqref{smasterequation1} in the direct-sum spin space and derive the steady-state solution.
Then by switching from the direct-sum spin space to the tensor-product spin space, we will analyze the spin population (polarization) in each domain 
and the quantum entanglement between the two domains. 

\subsection{Tensor-Product and Direct-Sum Spin Subspaces}
At initial time, the double domain system under consideration has a structure represented by Eq. \eqref{initialstate}, i.e. $
\big{|}\text{is}\big{\rangle} =|\uparrow\ldots\uparrow\rangle_{\text{A}}\otimes|\downarrow\ldots\downarrow\rangle_{\text{B}}$.
The initial state \eqref{initialstate} is fully symmetric in each domain but is not in the total spin system $D=D_{\text{A}}+D_{\text{B}}$.
Here we mean the symmetric state as a state which is fully invariant under the permutation between any two spins.

The total Hamiltonian \eqref{hamiltonian1} is described by the total spin $J^\alpha=J^\alpha_{\text{A}}+J^\alpha_{\text{B}}$ and satisfies $[\boldsymbol{J}^2,H]=[(J^x)^2+(J^y)^2+(J^z)^2,H]=0$, which means that 
the total spin angular momentum is conserved and $[\boldsymbol{J}_{\text{A(B)}}^2,H]=[(J_{\text{A(B)}}^x)^2+(J_{\text{A(B)}}^y)^2+(J_{\text{A(B)}}^z)^2,H]=0$, implying the conservation of the angular moment of each spin domain.
These conditions constraint the dynamics of the system.  
To capture this, we employ the direct-sum spin state representation.  
This allows us to largely reduce the Hilbert space to analyze the dynamics. Then, later we transform the state of interest to the composite picture (tensor-product representation) to evaluate the entanglement between the domains. 
In the direct-sum representation, we can easily identify which subspaces are relevant to the system dynamics.  The mechanism of the collective relaxation in this system then becomes clearly understood and the steady-state formula is simply calculated.

In preparation for spin state analysis, let us introduce the above two spin spaces and explain their relations.    
First, the total spin space is given by 
\begin{align}
V_{\text{tot}}=\cal{H}_{\text{A}} \otimes \cal{H}_{\text{B}}  \label{spinspace1},
\end{align}
where $\cal{H}_{\text{A}}$ and $\cal{H}_{\text{B}}$ are spin subspaces
whose dimensions are $2^{N_A}$ and $2^{N_B}$, respectively, giving the total dimension of $2^{N_A+N_B}$ for $V_{\text{tot}}$.  
From the spin angular momentum conservation $[\boldsymbol{J}_{\text{A(B)}}^2,H]=0$ and the symmetry of the initial state \eqref{initialstate}, 
the Hilbert space which describes the system dynamics is highly reduced from the full space $V_{\text{tot}}$. 
We will call it $V_{\text{rel}}$, and next, let us analyze its structure. 
We introduce the two subspaces $V^{\text{sym}}_{\text{A}}$ and $V^{\text{sym}}_{\text{B}}$ which are  symmetric with respect to $\boldsymbol{J}_{\text{A}}$ and $\boldsymbol{J}_{\text{B}}$, respectively.  
The subspace $V^{\text{sym}}_{\text{A(B)}}$ is spanned by the eigenstates $| m_{\text{A(B)}}\rangle_{\text{A(B)}}$ which satisfy
$\boldsymbol{J}_{\text{A(B)}}^2| m_{\text{A(B)}}\rangle_{\text{A(B)}}=j_{\text{A(B)}}(j_{\text{A(B)}}+1)| m_{\text{A(B)}}\rangle_{\text{A(B)}}$ and $J_{\text{A(B)}}^z| m_{\text{A(B)}}\rangle_{\text{A(B)}}=m_{\text{A(B)}}| m_{\text{A(B)}}\rangle_{\text{A(B)}}$. Here $j_{\text{A(B)}}=N_{\text{A(B)}}/2$ and $m_{\text{A(B)}}=j_{\text{A(B)}},j_{\text{A(B)}}-1,\ldots,-j_{\text{A(B)}},$ are quantum numbers. 
The initial state \eqref{initialstate} is described in the form $| m_{\text{A}}\rangle_{\text{A}} \otimes | m_{\text{B}}\rangle_{\text{B}}$ which are the basis vectors of the tensor-product subspace $V^{\text{sym}}_{\text{A}}\otimes V^{\text{sym}}_{\text{B}}.$ 
On the other hand, the total Hamiltonian \eqref{hamiltonian1} or the Lindblad operator in Eq. \eqref{smasterequation1} is described by the total spin operator $J^\alpha$.
This means that the initial state \eqref{initialstate} decays by  the total spin operator and the spin state is described in terms of the states in $V^{\text{sym}}_{\text{A}}\otimes V^{\text{sym}}_{\text{B}}$ for arbitrary time.
Therefore, the subspace $V_{\text{rel}}$ is identified with $V^{\text{sym}}_{\text{A}}\otimes V^{\text{sym}}_{\text{B}}.$ 
Furthermore, the spin domain $D_{\text{A(B)}}$ behaves as a collective spin $J_{\text{A(B)}}$ whose spin size is equal to $N_{\text{A(B)}}/2$ owing to this Hilbert-space identification.
The dimension of the subspace $V_{\text{rel}}$ is $(N_{\text{A}}+1)(N_{\text{B}}+1)$ which is sufficiently smaller than that of  $V_{\text{tot}}$. 
The focus on $V_{\text{rel}}$ makes the analysis of the system dynamics simple and effective.

Now we convert the $V_{\text{rel}}$ to the direct-sum representation by the spin-angular-momentum composition of $\boldsymbol{J}_{\text{A}}$ and $\boldsymbol{J}_{\text{B}}$, which is described as \cite{JJQMtxb} 
\begin{align}
V_{\text{rel}}&=V^{\text{sym}}_{\text{A}}\otimes V^{\text{sym}}_{\text{B}}\notag\\
&=V_{j_{\text{A}}+j_{\text{B}}}\oplus V_{j_{\text{A}}+j_{\text{B}}-1}\oplus V_{j} \oplus
\ldots V_{j_{\text{A}}-j_{\text{B}}},  \label{spinspace2}
\end{align}
where $V_j$  is the subspace spanned by the basis $\{ | j;m_j\rangle\rangle | -j \leq m_j \leq j  \}$ 
where $m_j$ is a quantum number (a half integer)  given as $J_z|j;m_j\rangle\rangle=m_j|j;m_j\rangle\rangle$. These basis vectors satisfy $\boldsymbol{J}^2 | j;m_j\rangle\rangle=j(j+1) | j;m_j\rangle\rangle$.
The largest subspace $V_{j_{\text{A}}+j_{\text{B}}}$ is spanned by the fully symmetric spin states which we just call it a symmetric subspace while the other subspaces as asymmetric subspaces. 

Finally, the eigenstates $|j;m_j \rangle\rangle$ in $V_j$ are related to the basis vectors $| m_{\text{A}}\rangle_{\text{A}}  \otimes | m_{\text{B}}\rangle_{\text{B}}$ ($\in V^{\text{sym}}_{\text{A}}\otimes V^{\text{sym}}_{\text{B}}$) via the Clebsh-Gordan coefficients $C^{jm}_{m_{\text{A}}m_{\text{B}}}=\langle\langle j;m_j |m_{\text{A}}\rangle_{\text{A}} \otimes |m_{\text{B}}\rangle_{\text{B}}$.

\subsection{Dynamics and Steady State}\label{DME1}
\begin{figure}[t] 
\includegraphics[width=0.3 \textwidth]{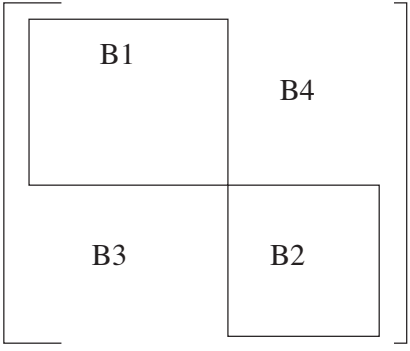}
\caption{The density matrix structure in the direct-sum spin space for  $N_{\text{A}}$=$N$ ($\geq1$) and $N_{\text{B}}=1.$
The diagonal blocks B$_1$ and B$_2$ are represented by the basis vectors $\boldsymbol{e}_1\sim\boldsymbol{e}_{N+2}$ and 
$\boldsymbol{e}_{N+3}\sim\boldsymbol{e}_{2(N+1)}$, respectively. The blocks B$_3$ and B$_4$ describe the off-diagonal parts. } 
\label{densitymatrixstructure} 
\end{figure}

We will now investigate the spin relaxations in the double domain system by solving the master equation \eqref{smasterequation1}
in the direct-sum spin space \eqref{spinspace2}. 
As a first step, we take a spin configuration  $N_{\text{A}}$=$N$ ($\geq1$) and $N_{\text{B}}=1$ with the initial state \eqref{initialstate} as the simplest case.
As the initial state has the populations only in the two subspaces $V_{j_{\text{I}}}$ and $V_{j_{\text{II}}}$ ($j_1=(N+1)/2, j_2=(N-1)/2$) and the $\boldsymbol{J}^2$ is a conserved observable, we only need these two subspaces to represent the dynamics.  The relevant Hilbert subspace is given by
\begin{align}
V_{\text{rel}}
=V_{j_{\text{I}}}\oplus V_{j_{\text{II}}}.  \label{spinspace3}
\end{align}
$V_{j_{\text{I}}}$ is the symmetric subspace whereas  $V_{j_{\text{II}}}$ an asymmetric subspace.  We illustrate the relevant space $V_{\text{rel}}$ in a matrix form in Fig. \ref{densitymatrixstructure}.
This property of the representation space is powerful both in analytical calculations and in numerical calculations.

We can solve the master equation  \eqref{smasterequation1} in the direct-sum spin space \eqref{spinspace3} by
deriving the equations of motion for the matrix elements of the density matrix $\rho(t)$.
 First, we will label the spin states $| j_{\text{I(II)}};m_{\text{I(II)}}\rrangle$ as 
 \begin{align}
&\boldsymbol{e}_1=\Big{|}j_{\text{I}};j_{\text{I}}\rrangle[\Big], \ldots,
 \boldsymbol{e}_{N+2}=\Big{|}j_{\text{I}};-j_{\text{I}}\rrangle[\Big], \notag\\
&\boldsymbol{e}_{N+3}=\Big{|}j_{\text{II}}; j_{\text{II}}\rrangle[\Big], \ldots, 
 \boldsymbol{e}_{2(N+1)}=\Big{|}j_{\text{II}};- j_{\text{II}}\rrangle[\Big].\label{basis1}
\end{align}
 Second, we will label the rows and columns of the density matrix $\rho$ using the basis vectors \eqref{basis1}. The matrix elements are obtained as
 \begin{align}
 &\rho_{\alpha_{\text{I}},\alpha_{\text{I}}^\prime}={}_{\text{I}}\llangle j_{\text{I}};m^z_{\alpha_{\text{I}}}| \rho  | j_{\text{I}};m^z_{\alpha^\prime_{\text{I}}}\rrangle_{\text{I}}, \notag\\
 & \rho_{\alpha_{\text{II}},\alpha_{\text{II}}^\prime}={}_{\text{I}}\llangle j_{\text{II}};m^z_{\alpha_{\text{II}}}| \rho  | j_{\text{II}};m^z_{\alpha^\prime_{\text{II}}}\rrangle_{\text{I}}, \notag\\
 &\rho_{\alpha_{\text{I}},\alpha_{\text{II}}}={}_{\text{I}}\llangle j_{\text{I}};m^z_{\alpha_{\text{I}}}| \rho  | j_{\text{II}};m^z_{\alpha_{\text{II}}}\rrangle_{\text{I}}. \label{Nvs1matrixelements}
 \end{align}
Here the indices $\alpha_{\text{I}},\alpha^\prime_{{\text{I}}}$ run from 1 to $N+2$ whereas $\alpha_{\text{II}},\alpha^\prime_{\text{II}}$ running from $N+3$ to $2N+2$. 
The values $m^{z}_{\alpha_{\text{I}}}$ and $m^{z}_{\alpha_{\text{II}}}$ are the eigenvalues corresponding to the eigenstates $\boldsymbol{e}_{\alpha_{\text{I}}}$ and $\boldsymbol{e}_{\alpha_{\text{II}}}$
in Eq. \eqref{basis1}, respectively.
The state $| j_{\text{I(II)}};m^z_{\alpha_{\text{I(II)}}}\rrangle_{\text{I}}$ is defined by $ | j_{\text{I(II)}};m^z_{\alpha_{\text{I(II)}}}\rrangle_{\text{I}}=\exp(i\omega J^zt)| j_{\text{I(II)}};m^z_{\alpha_{{\text{I(II)}}}}\rrangle.$
As presented in Fig. \ref{densitymatrixstructure}, the representation of density matrix $\rho$ in the direct-sum spin space  is described in terms of four blocks; A block B$_1$ is the symmetric part labeled by the basis vectors 
$\boldsymbol{e}_1\sim\boldsymbol{e}_{N+2}$ and the matrix elements here are given by $\rho_{\alpha_{\text{I}},\alpha_{\text{I}}^\prime}$ in Eq. \eqref{Nvs1matrixelements}.  
A block B$_2$ is the asymmetric part labeled by $\boldsymbol{e}_{N+3}\sim\boldsymbol{e}_{2(N+1)}$. The corresponding matrix elements are $\rho_{\alpha_{\text{II}},\alpha_{\text{II}}^\prime}$ in Eq. \eqref{Nvs1matrixelements}. 
 Blocks B$_3$ and B$_4$ are the cross terms between the symmetric and asymmetric parts. The matrix elements in the B$_3$ are given by in $\rho_{\alpha_{\text{I}},\alpha_{\text{II}}}$ in Eq. \eqref{Nvs1matrixelements} 
 and their Hermitian conjugates are equal to the matrix elements in the block B$_4$.
Third,  by multiplying $\llangle j_{\text{I(II)}};m^z_{\alpha_{{\text{I(II)}}}}|$ to the left hand side of Eq. \eqref{smasterequation1} while  $| j_{{\text{I(II)}}};m^z_{\alpha_{\text{I(II)}}}\rrangle$ to the right hand side of it,
we have the equations of motion for the matrix elements
\begin{widetext}
\begin{align}
\dot{\rho}_{\alpha_{\text{I}},\alpha^\prime_{\text{I}} }&=2\gamma \Bigg{[}
\left(j_{\text{I}}-m^{z}_{\alpha_{\text{I}}}  \right)\left(j_{\text{I}}+m^{z}_{\alpha_{\text{I}}} +1  \right)
\left(j_{\text{I}}-m^{z}_{\alpha^\prime_{\text{I}}}  \right)\left(j_{\text{I}}+m^{z}_{\alpha^\prime_{\text{I}}}+1  \right)
\Bigg{]}^{\frac{1}{2}}{\rho}_{\alpha_{\text{I}} -1,\alpha^\prime_{\text{I}} -1} \notag\\
&-\gamma \bigg{[}
\left(j_{\text{I}}+m^{z}_{\alpha_{\text{I}}}   \right)\left(j_{\text{I}}-m^{z}_{\alpha_{\text{I}}} +1  \right)
+\left(j_{\text{I}}+m^{z}_{\alpha^\prime_{\text{I}}} \right)\left(j_{\text{I}}-m^{z}_{\alpha^\prime_{\text{I}}} +1  \right)
\Bigg]{\rho}_{\alpha_{\text{I}}, \alpha^\prime_{\text{I}} }, 
\label{eom1}\\
\dot{\rho}_{\alpha_{\text{II}},\alpha^\prime_{\text{II}}} &=2\gamma \Bigg{[}
\left(j_{\text{II}}-m^{z}_{\alpha_{\text{II}}}  \right)\left(j_{\text{II}}+m^{z}_{\alpha_{\text{II}}} +1  \right)
\left(j_{\text{II}}-m^{z}_{\alpha^\prime_{\text{II}}}  \right)\left(j_{\text{II}}+m^{z}_{\alpha^\prime_{\text{II}}}+1  \right)
\Bigg{]}^{\frac{1}{2}}{\rho}_{\alpha_{\text{II}} -1,\alpha^\prime_{\text{II}} -1} \notag\\
&-\gamma \bigg{[}
\left(j_{\text{II}}+m^{z}_{\alpha_{\text{II}}}   \right)\left(j_{\text{II}}-m^{z}_{\alpha_{\text{II}}} +1  \right) 
+\left(j_{\text{II}}+m^{z}_{\alpha^\prime_{\text{II}}} \right)\left(j_{\text{II}}-m^{z}_{\alpha^\prime_{\text{II}}} +1  \right)
\Bigg]{\rho}_{\alpha_{\text{II}}, \alpha^\prime_{\text{II}} } , 
\label{eom2}\\
\dot{\rho}_{\alpha_{\text{I}},\alpha_{\text{II}}} &=2\gamma \Bigg{[}
\left(j_{\text{I}}-m^{z}_{\alpha_{\text{I}}}  \right)\left(j_{\text{I}}+m^{z}_{\alpha_{\text{I}}} +1  \right)
\left(j_{\text{II}}-m^{z}_{\alpha_{\text{II}}}  \right)\left(j_{\text{II}}+m^{z}_{\alpha_{\text{II}}}+1  \right)
\Bigg{]}^{\frac{1}{2}}{\rho}_{\alpha_{\text{I}} -1,\alpha_{\text{II}} -1} \notag\\
&-\gamma \bigg{[}
\left(j_{\text{I}}+m^{z}_{\alpha_{\text{I}}}   \right)\left(j_{\text{I}}-m^{z}_{\alpha_{\text{I}}} +1  \right)
+\left(j_{\text{II}}+m^{z}_{\alpha_{\text{II}}} \right)\left(j_{\text{II}}-m^{z}_{\alpha_{\text{II}}} +1  \right)
\Bigg]{\rho}_{\alpha_{\text{I}}, \alpha_{\text{II}} }. 
\label{eom3}
\end{align}
\end{widetext}

\begin{figure*}[t]
\includegraphics[width=0.7\textwidth]{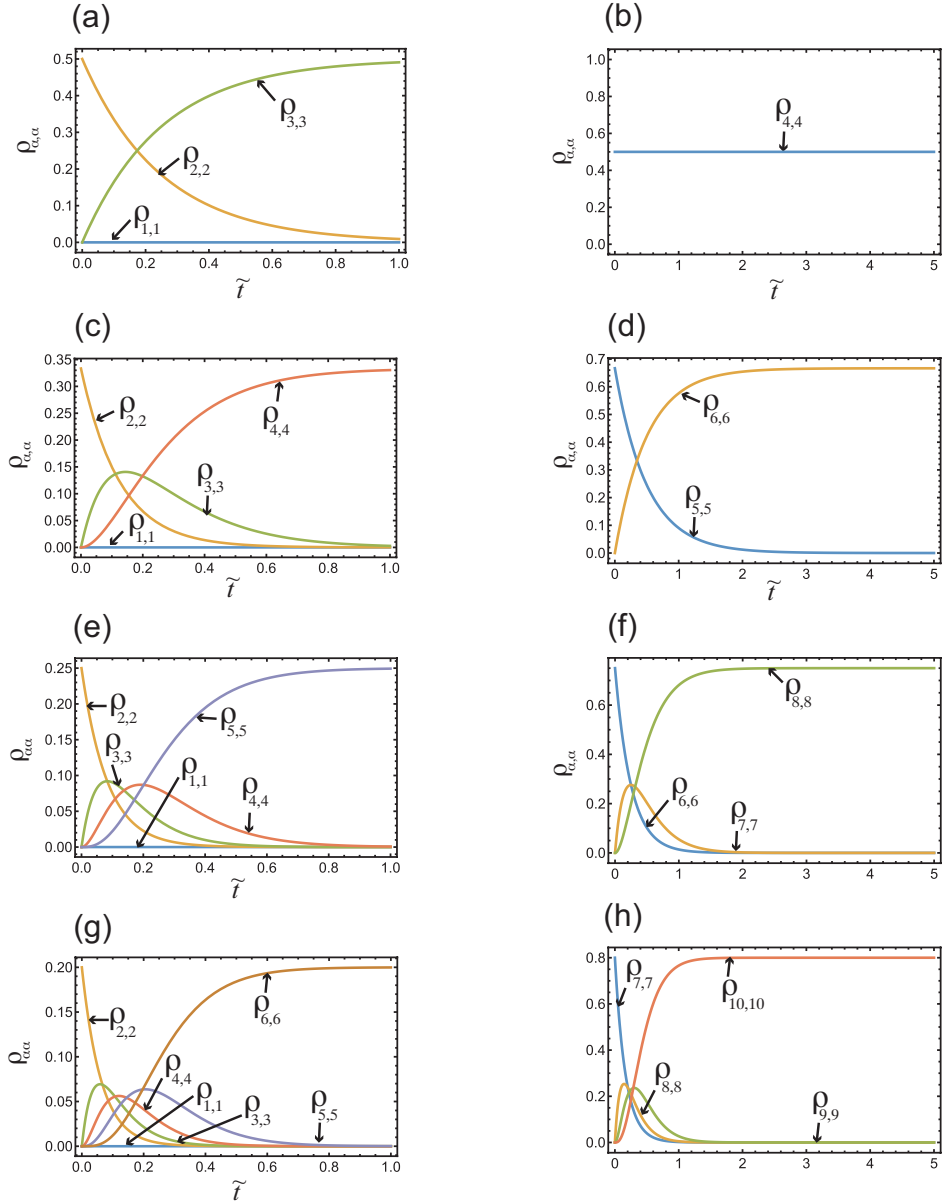}
\caption{The dynamics of the diagonal components of density matrix in the direct-sum spin space. The horizontal axis denotes the dimensionless time $\tilde{t}$.
(a), (c), (e), and (g) are the dynamics of the diagonal elements in the block B$_1$ while (b), (d), (f), and (h) are those in the block B$_2$ for  $N=1,2,3,$ and 4,  respectively. $N_{\text{B}}$ is fixed to one for all figures.
The only components which survive at the steady state are the end points of diagonal blocks: $\rho_{N+2,N+2}$ in the block B$_1$ and $\rho_{2N+2,2N+2}$ in the block B$_2$. $\rho_{N+2,N+2}$ converges to $1/(N+1)$ whereas $\rho_{2N+2,2N+2}$ to $N/(N+1)$.  }
\label{DMdiagonal}  
\end{figure*}

Equations \eqref{eom1}, \eqref{eom2}, and \eqref{eom3} are the equations of motion for the matrix elements in the blocks B$_1$, B$_2$, and B$_3$, respectively.
The equations of motion for the matrix elements in the block B$_4$ are obtained by taking the Hermitian conjugate of Eq. \eqref{eom3}. 
To derive the above equations we have used the relations $J^{\pm}J^{\mp}=\boldsymbol{J}^2-(J^z)^2\pm J^z$ and
$J^{\pm}_a|j_a,m_a \rrangle=\sqrt{j_a(j_a+1)-m_a(m_a\pm1)}|j_a,m_a\pm1 \rangle\rangle$ with $a=$I,II. \\ 
The initial state \eqref{initialstate} for this case is given by
\begin{align}
|\text{is}\rangle=\Big{|} \frac{N}{2}\Big{\rangle}_{\text{A}}  \otimes \Big{|} -\frac{1}{2}\Big{\rangle}_{\text{B}}. 
\label{initialstate2}
\end{align}
Now by using the relations \cite{angularmomentumtextbook} 
\begin{align}
\Big{|} j_{\text{I}};\frac{N-1}{2}\rrangle[\Big]&= \left(\frac{1}{N+1}\right)^{\frac{1}{2}} \Big{|} \frac{N}{2}\Big{\rangle}_{\text{A}}\otimes\Big{ | }-\frac{1}{2}\Big{\rangle}_{\text{B}}\notag\\
 &+\left(\frac{N}{N+1}\right)^{\frac{1}{2}}\Big{ | }\frac{N-2}{2}\Big{\rangle}_{\text{A}}\otimes \Big{|} \frac{1}{2}\Big{\rangle}_{\text{B}},  \notag\\
 \Big{ |} j_{\text{II}};\frac{N-1}{2}\rrangle[\Big]&=\left(\frac{N}{N+1}\right)^{\frac{1}{2}} \Big{|} \frac{N}{2}\Big{\rangle}_{\text{A}}\otimes\Big{ | }-\frac{1}{2}\Big{\rangle}_{\text{B}}\notag\\
 &-\left(\frac{1}{N+1}\right)^{\frac{1}{2}}\Big{ | }\frac{N-2}{2}\Big{\rangle}_{\text{A}}\otimes \Big{|} \frac{1}{2}\Big{\rangle}_{\text{B}}, \label{relations1}
 \end{align}
 the density matrix for the initial state \eqref{initialstate2} can be represented in the direct-sum spin space as
\begin{widetext}
\begin{align}
 \rho_{\text{is}}(N)&=\frac{1}{N+1}\Big{|} j_{\text{I}};\frac{N-1}{2}\rrangle[\Big]   \llangle[\Big] j_{\text{I}};\frac{N-1}{2}\Big{|}
 +\frac{N}{N+1}\Big{|}  j_{\text{II}},\frac{N-1}{2}\rrangle[\Big]    \llangle[\Big]  j_{\text{II}};\frac{N-1}{2}\Big{|}\notag\\
&+\frac{\sqrt{N}}{N+1}\bigg{(}\Big{|} j_{\text{I}};\frac{N-1}{2}\rrangle[\Big]   \llangle[\Big]  j_{\text{II}};\frac{N-1}{2}\Big{|}
+\Big{|}  j_{\text{II}};\frac{N-1}{2}\rrangle[\Big]    \llangle[\Big] j_{\text{I}};\frac{N-1}{2}\Big{|}\bigg{)}, \label{initialstate2-2}
 \end{align}
 \end{widetext}
   or the more compact form
\begin{align}
 &\left( \rho_{\text{is}}(N)\right)_{2,2}=\frac{1}{N+1}, \quad \left( \rho_{\text{is}}(N)\right)_{N+3,N+3}=\frac{N}{N+1},\notag\\
  &\left( \rho_{\text{is}}(N)\right)_{2,N+3}=\left( \rho_{\text{is}}(N)\right)_{N+3,2}=\frac{\sqrt{N}}{N+1}, \label{initialstate3}
\end{align}  
  with all the rest equal to zero. We will solve the Eqs. \eqref{eom1}-\eqref{eom3} under the initial conditions \eqref{initialstate3}.
Due to the factors appearing as $j_{\text{I,II}}$ and $m^z_{\alpha_{\text{I,II}}}$ in Eqs. \eqref{eom1}-\eqref{eom3},   
we can describe the effective dynamics of the matrix elements by two damping rates enhanced by $N.$  
This reflects that the double spin domain system exhibits the collective decay induced by the common reservoir.  
In the real systems, there are some effects which break this collective decay such as dephasing effects.
Even if the dephasing effects were included we still could observe this collective decay in this double domain systems as long as its timescales is comparable to that of the dephasing process.

  \begin{figure*}[t]
\includegraphics[width=0.8\textwidth]{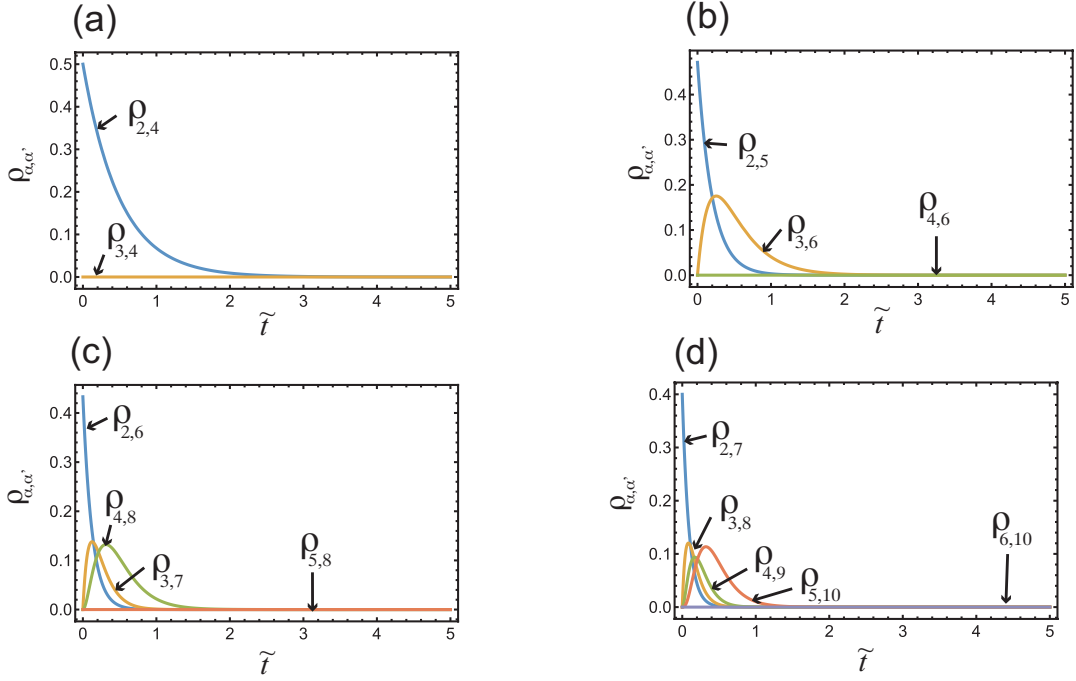}
\caption{The dynamics of the off-diagonal components of density matrix in the direct-sum spin space. The horizontal axis describes the dimensionless time $\tilde{t}$.  
(a), (b), (c), and (d) are the dynamics of the off-diagonal components in the block B$_3$ for $N=1,2,3,$ and 4, respectively.
All the  components vanish at the steady state. }   
\label{DMoffdiagonal}  
\end{figure*}
  
To see the dynamics of the matrix elements visually and what is occurring, we solve the  Eqs. \eqref{eom1}-\eqref{eom3}  for $N=1,2,3,$ and 4.
What we are particularly interested in is the dynamics of matrix elements which contributes to the relaxation of smaller spin $J^z_{\text{B}}$,
because as we see later this shows the negative-temperature state relaxation.
Thus, we analyze the dynamics of all the diagonal components as well as the off-diagonal elements contribute to the expectation values of $J^z_{\text{B}}$. For instance, 
in the case of $N=2$ the expectation $\langle J^z_{\text{B}}\rangle$ is described by $\langle J^z_{\text{B}} \rangle=
\frac{1}{6} \big{[}4\sqrt{2} \text{Re}( \rho_{2,5} + \rho_{3,6})
+ 3 \rho_ {1,1}+ \rho_{2,2} - \rho_{3,3} -3\rho_{4,4}- \rho_{5,5}+ \rho_{6,6}\big{]}.$ 
In Fig. \ref{DMdiagonal}, we present the time evolution of the diagonal components. The horizontal axis represents the dimensionless time defined by $\tilde{t}=\gamma t.$
  Figs. \ref{DMdiagonal} (a), (c), (e), and (g) plot the dynamics of the diagonal elements in the block B$_1$ whereas (b), (d), (f), and (h) display those in the block B$_2$ for $N=1,2,3,$ and 4, respectively.
  From these eight figures, what we see is that only the diagonal components $\rho_{N+2,N+2}$ and $\rho_{2N+2,2N+2}$, 
  which are the end points of the blocks B$_1$ and B$_2$, respectively,  survive at the steady state. The matrix element  $\rho_{N+2,N+2}$ converges to $1/(N+1)$ while $\rho_{2N+2,2N+2}$ to $N/(N+1)$.
  This indicates that in each block the upper components are going toward the end points with preserving the probability weight of the diagonal components given at the initial time:
  Denoting the density matrix for the initial and steady states as $\rho_{\text{is}}(N)$ and $\rho_{\text{ss}}(N)$, respectively, we see that 
in the block B$_1$ all the diagonal components except for the end point $\rho_{N+2,N+2}$  vanish such that 
$\left( \rho_{\text{is}}(N)\right)_{2,2}=\left( \rho_{\text{ss}}(N)\right)_{N+2,N+2}$. Similarly, in the block B$_2$ only the end point $\rho_{2N+2,2N+2}$ survives such that
$\left( \rho_{\text{is}}(N)\right)_{N+3,N+3}=\left( \rho_{\text{ss}}(N)\right)_{2N+2,2N+2}$. 
In contrast, in Fig. \ref{DMoffdiagonal} we have demonstrated the dynamics of off-diagonal components in the block B$_3$ which contributes to the expectations of $J^z_{\text{B}}$.
All these matrix elements vanish at the steady state. 
We have also presented the dynamics of the matrix elements $\left( \rho_{\text{ss}}(N)\right)_{N+2,2N+2}$, which are the end point of the block B$_3$. It is zero for the entire time. This is because at first from Eq. \eqref{eom3}, the equation of motion for $\left( \rho(N)\right)_{3,N+3}$ is represented by the linear differential equation with its initial value zero, which means that $\left( \rho(N)\right)_{3,N+3}$ is zero for the entire time. Then again from Eq. \eqref{eom3}, $\left( \rho_{\text{ss}}(N)\right)_{4,N+4},\ldots,\left( \rho_{\text{ss}}(N)\right)_{N+1,2N+1}$ and 
$\left( \rho_{\text{ss}}(N)\right)_{N+2,2N+2}$ are zero for any time by the same reason for $\left( \rho(N)\right)_{3,N+3}$. 
Thus, even $\left( \rho_{\text{ss}}(N)\right)_{N+2,N+2}$ and $\left( \rho_{\text{ss}}(N)\right)_{2N+2,2N+2}$ are finite, their cross components $\left( \rho_{\text{ss}}(N)\right)_{N+2,2N+2}$ and 
$\left( \rho_{\text{ss}}(N)\right)_{2N+2,N+2}$ vanish.  By using the same argument, we can verify that all the other off-diagonal elements remain zero under the time evolution.
As a result, the only terms which survive at the steady state are $\rho_{N+2,N+2}$ and $\rho_{2N+2,2N+2}$.

 From the above analysis, we can establish (see Appendix \ref{Nvs1steadystateproof} for details) that for any $N$ the density matrix for the steady state has the form
\begin{align}
\rho_{\text{ss}}(N)&=\sum_{i=\text{I}}^{\text{II}}p_i
\Big{|} j_i;-j_i\rrangle[\Big]    \llangle[\Big] j_i;-j_i\Big{|},
  \label{steadystatedm1}
 \end{align}
with $p_{\text{I}}=1/(N+1),p_{\text{II}}=N/(N+1)$. The steady state \eqref{steadystatedm1} can be represented in the matrix form as
\begin{equation} 
	{\rho}_{\mathrm{ss}} (N) = 
	\left (
		\begin{array}{cccc|ccc} 
		 0 & & & & & & \\
		 & 0 & & & \multicolumn{3}{c}{\multirow{3}{*}{\mbox{\smash{\huge{$0$}}}}} \\
		 & & \ddots &  & & & \\ 
		 & & & p_{\text{I}} & & & \\ \hline
		 & & & & 0 & & \\ 
		 \multicolumn{4}{c|}{\mbox{\smash{\huge{$0$}}}} & & \ddots & \\ 
		 & & & & & &  p_{\text{II}}
		\end{array}
	\right ).
\end{equation} 
Next, let us look at the structure of the steady state \eqref{steadystatedm1}. 
The first terms represents the ground state of the total spin because in this state all the spins align in downward. The probability weight to be in this state is given by $1/(N+1)$.
The second term describes the asymmetric state and includes the effect inherent to the double domain structure \eqref{initialstate} with its probability weight $N/(N+1)$. 
This effect becomes stronger as $N$ gets larger leading to the novel relaxation processes which cannot be realized in the single spin domain system.   

\begin{figure*}[t]
\includegraphics[width=0.9\textwidth]{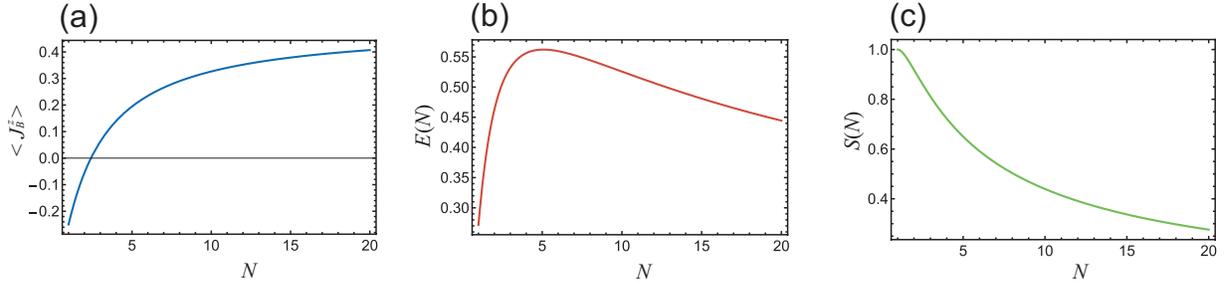}
\caption{ (a) Plot of $\langle J^z_{\text{B}}\rangle$ as a function of $N.$ The negative-temperature steady state starts to emerge from $N=3.$ 
(b) Plot of the amount of entanglement (logarithmic negativity) present as a function of $N.$ It takes a maximum at $N=5.$
(c)  Plot of the the von Neumann entropy as a function of $N.$ The steady state is maximally mixed at $N=1$ and becomes a pure state at $N\rightarrow\infty.$} 
\label{NTLNEE}  
\end{figure*}

\subsection{Negative-Temperature State Relaxation and Quantum Entanglement}\label{SSNTQE1}
Having established the form of the steady state, we will analyze the 
the spin polarization for each domain especially the polarization for the small domain $D_{\text{B}}$, and examine the quantum-entanglement creation between the two domains. 
To calculate these quantities, we rewrite the steady state \eqref{steadystatedm1} in the tensor-product space representation using the relations \cite{angularmomentumtextbook} 
\begin{align}
\Big{|} j_{\text{I}};-j_{\text{I}}\rrangle[\Big] &= \Big{|}-\frac{N}{2}\Big{\rangle}_{\text{A}}\otimes\Big{ | }-\frac{1}{2}\Big{\rangle}_{\text{B}}\notag\\
 \Big{ |} j_{\text{II}};-j_{\text{II}}\rrangle[\Big] &=-\sqrt{\frac{N}{N+1}} \Big{|} -\frac{N}{2}\Big{\rangle}_{\text{A}}\otimes\Big{ | }\frac{1}{2}\Big{\rangle}_{\text{B}}\notag\\
 &+\sqrt{\frac{1}{N+1}}\Big{ | }-\frac{N-2}{2}\Big{\rangle}_{\text{A}}\otimes \Big{|} -\frac{1}{2}\Big{\rangle}_{\text{B}}. \label{relations2}
 \end{align}
The steady-state density matrix in the tensor-product spin space can be expressed as
\begin{widetext}
\begin{align}
\rho_\text{ss}(N)&=\frac{1}{(N+1)}\Big{|} -\frac{N}{2}\Big{\rangle}_{\text{AA}}  \Big{\langle}-\frac{N}{2}\Big{|}\otimes \Big{|}- \frac{1}{2}\Big{\rangle} _{\text{BB}}   \Big{\langle}-\frac{1}{2}\Big{|}
+\frac{N^2}{(N+1)^2}\Big{|} -\frac{N}{2}\Big{\rangle} _{\text{AA}} \Big{\langle}-\frac{N}{2}\Big{|}\otimes \Big{|} \frac{1}{2}\Big{\rangle} _{\text{BB}}  \Big{\langle}\frac{1}{2}\Big{|}\notag\\
&+
\frac{N}{(N+1)^2}\Big{|} -\frac{N-2}{2}\Big{\rangle}_{\text{AA}} \Big{\langle}-\frac{N-2}{2}\Big{|}\otimes \Big{|} -\frac{1}{2}\Big{\rangle}_{\text{BB}} \Big{\langle}-\frac{1}{2}\Big{|}\notag\\
&-\frac{N^{3/2}}{(N+1)^2}\Big(\Big{|} -\frac{N-2}{2}\Big{\rangle} _{\text{AA}} \Big{\langle}-\frac{N}{2}\Big{|}\otimes \Big{|} -\frac{1}{2}\Big{\rangle} _{\text{BB}} \Big{\langle}\frac{1}{2}\Big{|}
+| -\frac{N}{2}\Big{\rangle} _{\text{AA}} \Big{\langle}-\frac{N-2}{2}\Big{|}\otimes \Big{|} \frac{1}{2}\Big{\rangle} _{\text{BB}} \Big{\langle}-\frac{1}{2}\Big{|}
\Big).
\label{steadystatedm2}
\end{align}
\end{widetext}

From the above equation we obtain the spin polarization in the domain $D_{\text{A}}$ at the steady state 
\begin{align}
\langle  J^z_{\text{A}} \rangle_{\text{ss}}(N)=\text{Tr}( J^z_{\text{A}}\rho_\text{ss}(N))=-\frac{N}{2}\cdot
\frac{(N+1)^2-2}{(N+1)^2},
\label{JzAexpvalue}
\end{align} 
while the spin polarization in the domain $D_{\text{B}}$ is
\begin{align}
\langle  J^z_{\text{B}} \rangle_{\text{ss}}(N)=\text{Tr}( J^z_{\text{B}}\rho_\text{ss}(N))=\frac{(N-1)^2-2}{2(N+1)^2}. \label{JzBexpvalue}
\end{align} 
We show the plot of $\langle  J^z_{\text{B}} \rangle_{\text{ss}}(N)$ in Fig. \ref{NTLNEE}  (a). 
We see that from $N=3,$ the $J^z_{\text{B}}$ becomes positive which means that the spin population in the excited state is larger than that in the ground state, i.e. {\it the negative-temperature state relaxation}. 
At $N\rightarrow\infty$, we have $\langle  J^z_{\text{B}} \rangle_{\text{ss}}\rightarrow1/2$ which means that $D_{\text{B}}$ is completely excited while 
$\langle  J^z_{\text{A}} \rangle_{\text{ss}}\rightarrow -N/2$ indicating that the larger spin domain $D_{\text{A}}$ is in the ground state.
The mechanism of the negative temperature relaxation is clearly understood from the density matrix \eqref{steadystatedm1}.
The first term describes the ground state in the symmetric space. In this subspace, initially the spin state is prepared in the second highest energy level $\boldsymbol{e}_2$\ in Eq. \eqref{basis1}
and decays to the state $\boldsymbol{e}_{N+2}$\ in Eq. \eqref{basis1}.
The second term in Eq. \eqref{steadystatedm1} represents the ground state in the asymmetric subspace. 
Initially, the spin state in this subspace is prepared in the highest energy level $\boldsymbol{e}_{N+3}$\ in Eq. \eqref{basis1}
and decays to the state $\boldsymbol{e}_{2N+2}$\ in Eq. \eqref{basis1}.
This process gives the excitation to the double spin domain so that $J^z_{\text{B}}$ obtains the positive-polarization contribution.
As mentioned previously, we see from Eq. \eqref{steadystatedm1} that the effect of the first term becomes smaller while that from the second term gets bigger as $N$ increases. 
Therefore, $J^z_{\text{B}}$ relaxes to the negative-temperature state and its effective temperature becomes lower as $N$ increases.

Next let us examine the quantum-entanglement creation between the two domains. From Eq.  \eqref{steadystatedm2} we see that the terms in the first and second lines are written in a form
$\sum_k w_k (\rho^{A}_k \otimes\rho^{B}_k)$ ($w_k\geq0$, $\sum_k w_k=1$) which is an expression for the density matrix of a quantum state in a separable state. 
The density matrix \eqref{steadystatedm2} is represented by this separable-state part and the additional terms which are written in the third line.
Therefore, we readily see that the quantum entanglement is generated between the two domains at the steady state, namely,  {\it the reservoir-assisted quantum entanglement}.  
The quantum entanglement generated by the common reservoir were also found in the different contexts,
for instance,  two-qubit systems \cite{Etwoqubitspra2002,Etwoqubitsprl2002,Etwoqubitspra2006,Etwochargequbitsprb2008,Etwoqubitsprl2008pra2009}, two harmonic-oscillator system \cite{EtwooscillatorsjphysA2008}, 
and quantum entanglement between two ions or atoms in a single ionic (atomic) ensemble (for other related topics of reservoir-assisted quantum entanglement, see for instance \cite{IOPreviewopendynamicsentanglement} and references within). Here we have found the reservoir-assisted quantum entanglement between the two spin domains as a consequence of the collective spin decay.

Let us evaluate the quantum entanglement between the two spin domains. Here we use the logarithmic negativity \cite{PlenioPRL2005}
\begin{align}
E(\rho)=\log_2 || \rho^{\Gamma_{\text{A}}}  ||_1,
\label{negativity}
\end{align} 
where $\Gamma_{\text{A}}$ denotes the partial transposition with respect to the subsystem A, 
and the trace norm $|| X ||_1$ is defined by $|| X ||_1=\text{Tr}|X|=\text{Tr}\sqrt{X^\dagger X}.$

First by taking the partial transpose to the density matrix \eqref{steadystatedm2}, we have
 \begin{widetext}
 \begin{align}
(\rho_\text{ss})^{\Gamma_{J_{\text{A}}}}(N)&=\frac{1}{(N+1)}\Big{|} -\frac{N}{2}\Big{\rangle}  _{\text{AA}}    \Big{\langle}   -\frac{N}{2}\Big{|}\otimes \Big{|}- \frac{1}{2}\Big{\rangle} _{\text{BB}} \Big{\langle}-\frac{1}{2}\Big{|}
+\frac{N^2}{(N+1)^2}\Big{|} -\frac{N}{2}\Big{\rangle} _{\text{AA}}   \Big{\langle}-\frac{N}{2}\Big{|}\otimes \Big{|} \frac{1}{2}\Big{\rangle} _{\text{BB}}  \Big{\langle}\frac{1}{2}\Big{|}\notag\\
&+
\frac{N}{(N+1)^2}\Big{|} -\frac{N-2}{2}\Big{\rangle} _{\text{AA}}  \Big{\langle}-\frac{N-2}{2}\Big{|}
\otimes \Big{|} -\frac{1}{2}\Big{\rangle} _{\text{BB}}  \Big{\langle}-\frac{1}{2}\Big{|}\notag\\
&-\frac{N^{3/2}}{(N+1)^2}\Big(\Big{|} -\frac{N-2}{2}\Big{\rangle} _{\text{AA}}  \Big{\langle}-\frac{N}{2}\Big{|}\otimes \Big{|} \frac{1}{2}\Big{\rangle} _{\text{BB}}  \Big{\langle}-\frac{1}{2}\Big{|}
+\Big{|} -\frac{N}{2}\Big{\rangle}_{\text{AA}}   \Big{\langle}-\frac{N-2}{2}\Big{|}\otimes \Big{|} -\frac{1}{2}\Big{\rangle} _{\text{BB}}  \Big{\langle}\frac{1}{2}\Big{|}
\Big).
\label{partialtsteadystatedm1}
\end{align}
\end{widetext}
We note here that $(\rho_\text{ss})^{\Gamma_{J_{\text{A}}}}(N)=(\rho_\text{ss})^{\Gamma_{J_{\text{B}}}}(N).$ 
By deriving the eigenvalues of $\rho^{\Gamma_{J_{\text{A}}}}_\text{ss}(N)$ (or $\rho^{\Gamma_{J_{\text{B}}}}_\text{ss}(N)$) the logarithmic negativity for the matrix \eqref{partialtsteadystatedm1} is given by
\begin{align}
E\big{[}(\rho_\text{ss})^{\Gamma_{J_{\text{A}}}}(N)\big{]}=\log_2 \left[ 
\frac{\sqrt{4N^3+(N+1)^2}+N^2+N}{(N+1)^2}
\right].
\label{negativity2}
\end{align} 
We present the logarithmic negativity \eqref{negativity2} in Fig. \ref{NTLNEE}  (b). 
It takes a maximum at $N=5$ and its value is around 0.56.  By comparing with the logarithmic negativities for the Bell states, which is equal to one,
we see that the two domains are quite entangled at this maximum point. The logarithmic negativity \eqref{negativity2} becomes zero as $N\rightarrow\infty.$
This can be easily understood from Eq. \eqref{steadystatedm2} because in this limit only the second term survives, which means that the steady state is in the separable state
$| -\frac{N}{2}\rangle\otimes | \frac{1}{2}\rangle$.

Finally, let us discuss how pure the steady state \eqref{steadystatedm2} is. We evaluate its purity by the von Neumann entropy defined by  
\begin{align}
S\Big[(\rho_{\text{ss}})(N)\Big]=-\rho_{\text{ss}}\log_2\rho_{\text{ss}}(N).
\label{entanglemententropy}
\end{align} 
From the eigenvalues of the steady state \eqref{steadystatedm2}, the von Neumann entropy becomes
\begin{align}
S\Big[(\rho_{\text{ss}})(N)\Big]&=-\frac{1}{N+1}\Big( \log_2\frac{1}{N+1}
+N\log_2\frac{N}{N+1} \Big).
\label{entanglemententropy2}
\end{align} 
We plot this as a function of $N$ in Fig.  \ref{NTLNEE} (c). 
The steady state \eqref{steadystatedm2} is maximally mixed at $N=1$ and the entropy takes one, and then it decreases as $N$ increases.
At $N\to\infty$, the entropy becomes zero which is consistent with the above argument for the quantum entanglement because the steady state \eqref{steadystatedm2} becomes the pure state in this limit.

The negative-temperature state relaxation \eqref{JzBexpvalue} and the reservoir-assisted quantum entanglement \eqref{negativity2} are the collective spin phenomena intrinsic to the double domain system driven by the common reservoir.
To see this clearly, let us compare the dynamics in a double spin system where each domain is individually coupling to a reservoir. 
Such dynamics is described by the Hamiltonian like Eq. \eqref{hamiltonian1} except the last interaction term is modified as 
$\hbar g_{\text{A}} \left(J^+_{\text{A}}R_{\text{A}}+J^-_{\text{A}}R_{\text{A}}^\dagger\right)/2+\hbar g_{\text{B}} \left(J^+_{\text{B}}R_{\text{B}}+J^-_{\text{B}}R_{\text{B}}^\dagger\right)/2.$
Each spin domain relaxes to its ground state and the steady state is a separable state in terms of the ground state of the first domain and that of the second domain. 
Therefore, both the negative-temperature relaxation and the reservoir-assisted entanglement are not realized in this case.

\section{Generalization to Larger Spin Systems }\label{generalization}
In this section, we will present the discussion for the spin configuration for $N_{\text{B}}\geq2$ (or the size of spin domain B larger than one).
First, we demonstrate the analysis in the case of $N_{\text{B}}=2$ by using the same argument which we did in Sec. \ref{DMS}. 
Then by comparing the results for the steady state in the cases of $N_{\text{B}}=1,2$, we will conjecture the steady-state solution for general $N_{\text{B}}$.

\begin{figure}[b] 
\includegraphics[width=0.35 \textwidth]{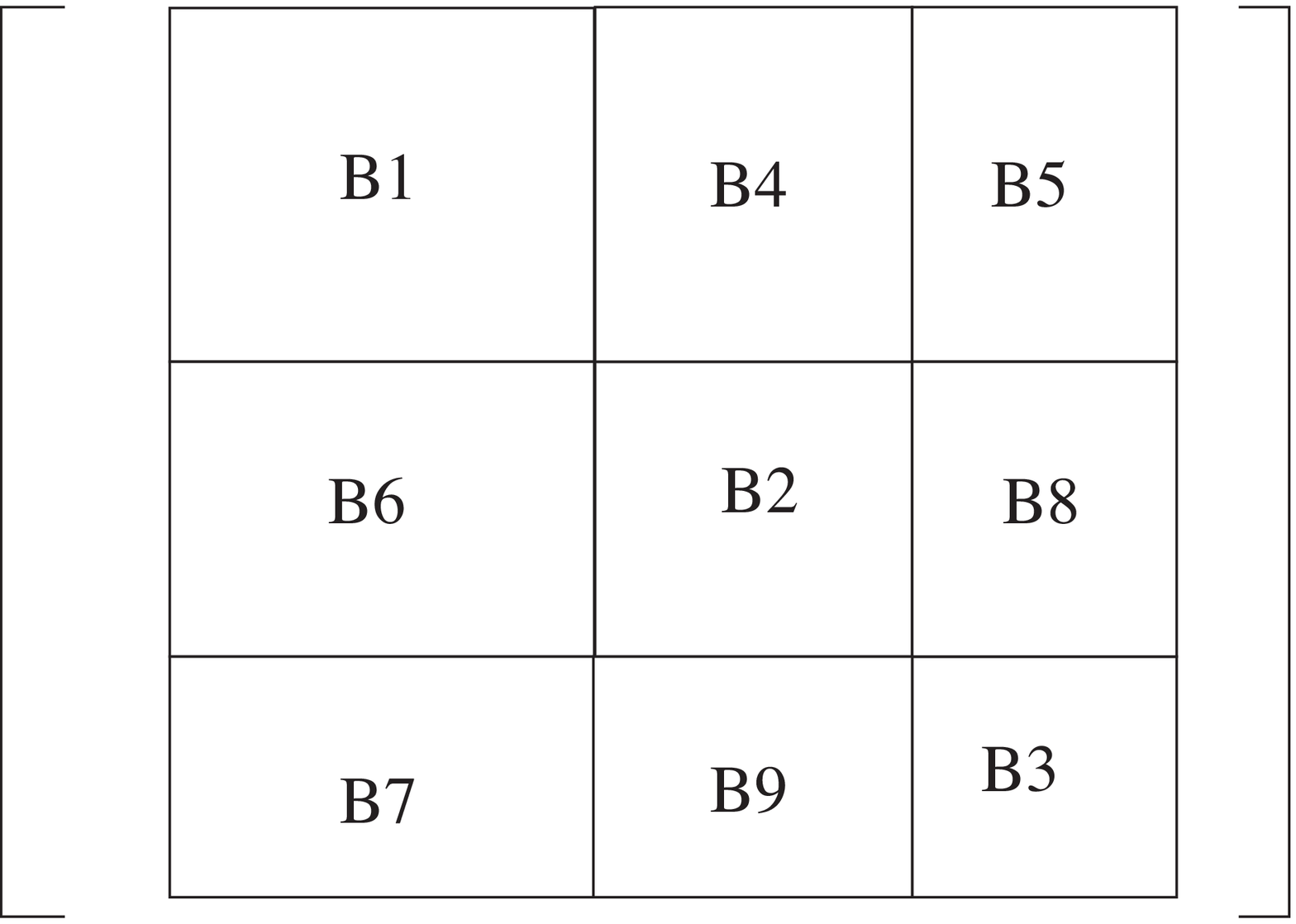}
\caption{The density matrix structure for represented by the direct-sum spin space. There are nine sub blocks and diagonal parts are the blocks B$_1$, B$_2$, and B$_3$.   } 
\label{densitymatrixstructure2} 
\end{figure}

The tensor-product spin space which describes the system dynamics is spanned by the eigenstates $| m_{\text{A}}\rangle_{\text{A}}\otimes | m_{\text{B}}\rangle_{\text{B}}$ with
$m_{\text{A}}=N/2,\ldots,-N/2$ and $m_{\text{B}}=1,0,-1$.
On the other hand, the corresponding direct-sum spin (symmetric-asymmetric) space has a structure
\begin{align}
V_{\text{rel}}=V_{j_1}\oplus V_{j_2}\oplus V_{j_3},  \label{modspinspace3}
\end{align}
where $j_1=(N/2)+1,j_2=N/2,$ and $j_3=(N/2)-1$.  Again, $V_{j_1}$, $V_{j_2}$, and $V_{j_3}$ are defined to accommodate the initial state.
The basis vectors which span the Hilbert space \eqref{modspinspace3}  are
\begin{align}
&\boldsymbol{e}_1=\Big{|} j_1; j_1\rrangle[\Big] , \ldots, \quad
 \boldsymbol{e}_{N+3}=\Big{|} j_1; -j_1\rrangle[\Big] , \notag\\
&\boldsymbol{e}_{N+4}=\Big{|} j_2; j_2\rrangle[\Big] , \ldots,\quad
 \boldsymbol{e}_{2(N+2)}=\Big{|}j_2; -j_2\rrangle[\Big] ,\notag\\
&\boldsymbol{e}_{2N+5}=\Big{|}j_3; j_3\rrangle[\Big] , \ldots,\quad
 \boldsymbol{e}_{3(N+1)}=\Big{|} j_3; -j_3\rrangle[\Big] .
\label{modbasis2}
\end{align}
The subspaces $V_{j_1}$, $V_{j_2}$, and $V_{j_3}$
are spanned by the eigenstates $\boldsymbol{e}_1\sim \boldsymbol{e}_{N+3},$ $\boldsymbol{e}_{N+4}\sim \boldsymbol{e}_{2(N+2)},$ and 
$\boldsymbol{e}_{2N+5}\sim \boldsymbol{e}_{3(N+1)},$ respectively. The subspace $V_{j_1}$ is the symmetric subspace.
The density matrix structure is represented by 9 blocks as depicted in Fig. \ref{densitymatrixstructure2}.
Blocks B$_1$, B$_2$, and B$_3$ are the diagonal parts constructed by the eigenvectors $\boldsymbol{e}_1\sim \boldsymbol{e}_{N+3},$ 
$\boldsymbol{e}_{N+4}\sim \boldsymbol{e}_{2(N+2)},\ldots$ and $\boldsymbol{e}_{2N+5}\sim \boldsymbol{e}_{3(N+1)},$ respectively.
The other blocks B$_4\sim$ B$_9$ are the off-diagonal parts; for instance, in the block B$_4$ the row is labeled by  $\boldsymbol{e}_1\sim \boldsymbol{e}_{N+2}$ whereas 
the column by $\boldsymbol{e}_{N+3}\sim \boldsymbol{e}_{2(N+2)}$.

Next, we derive the equations of motion for the matrix elements represented by the direct-sum spin space \eqref{modspinspace3}.
From our master equation \eqref{smasterequation1}, we have 
\begin{widetext}
\begin{align}
\dot{\rho}_{\alpha_{i},\alpha^\prime_{i}} &=2\gamma \Bigg{[}
\left(j_i-m^{z}_{\alpha_i}  \right)\left(j_i+m^{z}_{\alpha_i} +1  \right)
\left(j_i-m^{z}_{\alpha^\prime_i}  \right)\left(j_i+m^{z}_{\alpha^\prime_i}+1  \right)
\Bigg{]}^{\frac{1}{2}}{\rho}_{\alpha_i -1,\alpha^\prime_i -1} \notag\\
&-\gamma \bigg{[}
\left(j_i+m^{z}_{\alpha_i}   \right)\left(j_i-m^{z}_{\alpha_i} +1  \right)
+\left(j_i+m^{z}_{\alpha^\prime_i} \right)\left(j_i-m^{z}_{\alpha^\prime_i} +1  \right)
\Bigg]{\rho}_{\alpha_i, \alpha^\prime_i }, 
\label{modeom4}\\
\dot{\rho}_{\alpha_i,\alpha_{l}} &=2\gamma \Bigg{[}
\left(j_i-m^{z}_{\alpha_i}  \right)\left(j_i+m^{z}_{\alpha_i} +1  \right) 
\left(j_l-m^{z}_{\alpha_l}  \right)\left(j_l+m^{z}_{\alpha_l}+1  \right)
\Bigg{]}^{\frac{1}{2}}{\rho}_{\alpha_i -1,\alpha_l -1} \notag\\
&-\gamma \bigg{[}
\left(j_i+m^{z}_{\alpha_i}   \right)\left(j_i-m^{z}_{\alpha_i} +1  \right)
+\left(j_l+m^{z}_{\alpha_l} \right)\left(j_l-m^{z}_{\alpha_l} +1  \right)
\Bigg]{\rho}_{\alpha_i, \alpha_l },  \quad (i\neq l)
\label{modeom6}
\end{align}
\end{widetext}
with $i,l=1,2,3.$
The indices $\alpha_1,\alpha^\prime_{1}$ runs from 1 to $N+3$, 
whereas $\alpha_2,\alpha^\prime_{2}$ running from $N+4$ to $2(N+2)$, and  
$\alpha_3,\alpha^\prime_{3}$ from $2N+5$ to $3(N+1)$.
The value $m^{z}_{\alpha_i}$ is the eigenvalue of the eigenstate $\boldsymbol{e}_{\alpha_i}$ with respect to $J^z.$
We will solve the equations of motion  \eqref{modeom4} and \eqref{modeom6} under the initial condition 
\begin{align}
|\text{in}\rangle=\Big{|} \frac{N}{2}\Big{\rangle}_{\text{A}}  \otimes \Big{|} -1 \Big{\rangle}_{\text{B}}.
\label{initialstate4}
\end{align}
In the direct-sum spin space the initial state  \eqref{initialstate4} is expressed as \cite{angularmomentumtextbook} 

\begin{align}
|\text{in}\rangle&=\sqrt{\frac{2}{(N+1)(N+2)} }\Big{|}j_1; \frac{N}{2}-1\rrangle[\Big] \notag\\ 
&+\sqrt{\frac{2}{N+2} }\Big{|}j_2; \frac{N}{2}-1\rrangle[\Big] +\sqrt{\frac{N-1}{N+1} }\Big{|}j_3; \frac{N}{2}-1\rrangle[\Big] ,
\label{initialstate5}
\end{align}
or
\begin{align}
 &\left( \rho_{\text{is}}(N)\right)_{3,3}=\frac{2}{(N+1)(N+2)}, \notag\\
 & \left( \rho_{\text{is}}(N)\right)_{N+5,N+5}=\frac{2}{N+2},\notag\\
  &\left( \rho_{\text{is}}(N)\right)_{2N+5,2N+5}=\frac{N-1}{N+1},\notag\\
  &\left( \rho_{\text{is}}(N)\right)_{3,N+5}=\left( \rho_{\text{is}}(N)\right)_{N+5,3}=\frac{2}{N+2}\sqrt{\frac{1}{N+1}}, \notag\\
  &\left( \rho_{\text{is}}(N)\right)_{3,2N+5}=\left( \rho_{\text{is}}(N)\right)_{2N+5,3}\notag\\
  &=\frac{1}{N+1}\sqrt{\frac{2(N-1)}{N+2}}, \notag\\
   &\left( \rho_{\text{is}}(N)\right)_{N+5,2N+5}=\left( \rho_{\text{is}}(N)\right)_{2N+5,N+5}\notag\\
   &=\sqrt{\frac{2(N-1)}{(N+1)(N+2)}}, \notag\\
\label{initialstate6}
\end{align}
and the rest of components are zero. As in the case of $N_{\text{A}}=N,N_{\text{B}}=1,$ 
the two effective damping rates are enhanced as $N$ increases, indicating the collective decay.

We derive the steady-state density matrix.
 First for the matrix elements in the block B$_1$, from the initial condition \eqref{initialstate6}  we obtain $\left(\rho\right)_{1,1}(N,t)=\left(\rho\right)_{2,2}(N,t)=0.$
Then subsequently, we have
\begin{align}
\left(\rho\right)_{3,3}(N,t)&=\frac{2}{(N+1)(N+2)}\exp\Big{(} -6N\gamma t \Big{)}. \label{rho33solution}
 \end{align}
At the steady state, $\left(\rho\right)_{3,3}$ is zero, and subsequentially, we have $\left(\rho_{\text{ss}} \right)_{4,4}=\ldots=\left(\rho_{\text{ss}}\right)_{N+1,N+1}=0.$ 
Such argument can be exactly applied to the diagonal matrix elements in the blocks B$_2$ and B$_3$. Thus, the only elements which survive at the steady state are the end points of the blocks B$_1$ B$_2$, and B$_3$.
We have  $\left(\rho_{\text{ss}}\right)_{N+3,N+3}=p_1,\left(\rho_{\text{ss}}\right)_{2(N+2),2(N+2)}=p_2,\left(\rho_{\text{ss}}\right)_{3(N+1),3(N+1)}=p_3,$ where $p_1,p_2,p_3$ are the finite constants satisfying
$p_1+p_2+p_3=1.$ For off-diagonal elements, whether they have finite values or not at the initial state, they become zero at the steady state.   
Therefore, by considering that the spin subspaces $V_{j_1}$, $V_{j_2}$, and $V_{j_3}$ are orthogonal to each other, 
the constants $p_1,p_2,p_3$ must satisfy $p_1=\left( \rho_{\text{is}}(N)\right)_{3,3},p_2=\left( \rho_{\text{is}}(N)\right)_{N+5,N+5},p_3=\left( \rho_{\text{is}}(N)\right)_{2N+5,2N+5}.$ 
As a result, the density matrix at steady state in the direct-sum space representation has a form
\begin{align}
\rho_{\text{ss}}(N)&=\sum_{i=1}^3 p_i     \Big{|} j_i;-j_i\rrangle[\Big]   \llangle[\Big] j_i;-j_i\Big{|}, 
\label{steadystatedm3}
 \end{align}
 where $p_1=2/(N+1)(N+2),p_2=2/(N+2),p_3=(N-1)/(N+1).$
 In a matrix form, the steady state \eqref{steadystatedm3} is represented as 
\begin{equation} 
	{\rho}_{\mathrm{ss}} (N) =
	\left (
		\begin{array}{ccc|ccc|ccc} 
		 0 & & & & & & & & \\ 
		 & \ddots & & \multicolumn{3}{c|}{\mbox{\smash{\huge{$0$}}}} & \multicolumn{3}{c}{\mbox{\smash{\huge{$0$}}}} \\ 
		 & & p_1 & & & & & & \\ \hline
		 & & & 0 & & & & & \\ 
		 \multicolumn{3}{c|}{\mbox{\smash{\huge{$0$}}}} & & \ddots & & \multicolumn{3}{c}{\mbox{\smash{\huge{$0$}}}} \\ 
		 & & & & & p_2  & & & \\ \hline 
		 & & & & & & 0 & & \\ 
		 \multicolumn{3}{c|}{\mbox{\smash{\huge{$0$}}}} & \multicolumn{3}{c|}{\mbox{\smash{\huge{$0$}}}} & & \ddots & \\  
		 & & & & & & & & p_3
		\end{array}
	\right ).
\end{equation} 
Then from the relations \cite{angularmomentumtextbook} 
\begin{widetext}
 \begin{align}
\Big{|}j_1; -j_1\rrangle[\Big] &=\Big{|} -\frac{N}{2}\Big{\rangle}_{\text{A}}  \otimes \Big{|} -1\Big{\rangle}_{\text{B}},\notag\\
\Big{|}j_2; -j_2\rrangle[\Big] &=-\sqrt{\frac{N}{N+2}}\Big{|} -\frac{N}{2}\Big{\rangle}_{\text{A}}  \otimes \Big{|} 0 \Big{\rangle}_{\text{B}}+\sqrt{\frac{2}{N+2}}\Big{|} -\frac{N}{2}+1  \Big{\rangle}_{\text{A}}  \otimes \Big{|} -1\Big{\rangle}_{\text{B}}, \notag\\
\Big{|}j_3; -j_3\rrangle[\Big] &=  \sqrt{\frac{N-1}{N+1}}\Big{|} -\frac{N}{2}\Big{\rangle}_{\text{A}}  \otimes \Big{|} 1 \Big{\rangle}_{\text{B}}- \sqrt{\frac{2(N-1)}{N(N+1)}}\Big{|} 
-\frac{N}{2}+1\Big{\rangle}_{\text{A}}  \otimes \Big{|} 0 \Big{\rangle}_{\text{B}}
+\sqrt{\frac{2}{N(N+1)}} \Big{|} -\frac{N}{2}+2\Big{\rangle}_{\text{A}}  \otimes \Big{|} -1\Big{\rangle}_{\text{B}},
\end{align}
\end{widetext}
the steady-state density matrix can be represented in the tensor-product space as
\begin{widetext}
\begin{align}
\rho_\text{ss}(N)&=\frac{2}{(N+1)(N+2)}\Big{|} -\frac{N}{2}\Big{\rangle}_{\text{AA}} \Big{\langle}-\frac{N}{2}\Big{|}\otimes \Big{|}- 1\Big{\rangle} _{\text{BB}} \Big{\langle}-1\Big{|}
+\frac{2N}{(N+2)^2}\Big{|} -\frac{N}{2}\Big{\rangle}_{\text{AA}}  \Big{\langle}-\frac{N}{2}\Big{|}\otimes \Big{|} 0 \Big{\rangle} _{\text{BB}} \Big{\langle}0\Big{|}\notag\\
& +\frac{4}{(N+2)^2}\Big{|} -\frac{N}{2}+1\Big{\rangle}  _{\text{AA}} \Big{\langle}-\frac{N}{2}+1\Big{|}\otimes \Big{|} -1\Big{\rangle} _{\text{BB}} \Big{\langle} -1 \Big{|}
+\frac{(N-1)^2}{(N+1)^2}\Big{|} -\frac{N}{2}\Big{\rangle} _{\text{AA}} \Big{\langle}-\frac{N}{2}\Big{|}\otimes \Big{|} 1\Big{\rangle} _{\text{BB}} \Big{\langle}1\Big{|}\notag\\
& +\frac{2(N-1)^2}{N(N+1)^2}\Big{|} -\frac{N}{2}+1\Big{\rangle} _{\text{AA}} \Big{\langle}-\frac{N}{2}+1\Big{|}\otimes \Big{|} 0\Big{\rangle} _{\text{BB}} \Big{\langle} 0 \Big{|}
+\frac{2(N-1)}{N(N+1)^2}\Big{|} -\frac{N}{2}+2\Big{\rangle} _{\text{AA}} \Big{\langle}-\frac{N}{2}+2\Big{|}\otimes \Big{|} -1\Big{\rangle} _{\text{BB}} \Big{\langle}-1\Big{|}\notag\\
&-\frac{2\sqrt{2N}}{(N+2)^2}\Big(| -\frac{N}{2}\Big{\rangle} _{\text{AA}} \Big{\langle}-\frac{N}{2}+1\Big{|}\otimes \Big{|} 0\Big{\rangle} _{\text{BB}} \Big{\langle}-1\Big{|}
+\Big{|} -\frac{N}{2}+1\Big{\rangle} _{\text{AA}} \Big{\langle}-\frac{N}{2}\Big{|}\otimes \Big{|} -1 \Big{\rangle} _{\text{BB}} \Big{\langle} 0\Big{|}
\Big)\notag\\
&-\frac{(N-1)^2}{(N+1)^2}\sqrt{\frac{2}{N}}\Big(| -\frac{N}{2}\Big{\rangle} _{\text{AA}} \Big{\langle}-\frac{N}{2}+1\Big{|}\otimes \Big{|} 1\Big{\rangle} _{\text{BB}} \Big{\langle}0\Big{|}
+\Big{|} -\frac{N}{2}+1\Big{\rangle} _{\text{AA}} \Big{\langle}-\frac{N}{2}\Big{|}\otimes \Big{|} 0 \Big{\rangle} _{\text{BB}} \Big{\langle} 1\Big{|}
\Big)\notag\\
&+\sqrt{\frac{2(N-1)^3}{N(N+1)^4}}\Big(| -\frac{N}{2}\Big{\rangle} _{\text{AA}} \Big{\langle}-\frac{N}{2}+2\Big{|}\otimes \Big{|} 1\Big{\rangle} _{\text{BB}} \Big{\langle}-1\Big{|}
+\Big{|} -\frac{N}{2}+2\Big{\rangle} _{\text{AA}} \Big{\langle}-\frac{N}{2}\Big{|}\otimes \Big{|} -1 \Big{\rangle} _{\text{BB}} \Big{\langle} 1\Big{|}
\Big)\notag\\
&-\frac{2\sqrt{(N-1)^3}}{N(N+1)^2} \Big(| -\frac{N}{2}+1\Big{\rangle} _{\text{AA}} \Big{\langle}-\frac{N}{2}+2\Big{|}\otimes \Big{|} 0\Big{\rangle} _{\text{BB}}  \Big{\langle}-1\Big{|}
+\Big{|} -\frac{N}{2}+2\Big{\rangle} _{\text{AA}} \Big{\langle}-\frac{N}{2}+1\Big{|}\otimes \Big{|} -1 \Big{\rangle} _{\text{BB}} \Big{\langle} 0 \Big{|}
\Big).
\label{steadystatedm4}
\end{align}
\end{widetext}
The polarization of two domains are
\begin{align}
&\langle  J^z_{\text{A}} \rangle_{\text{ss}}(N)=-\frac{N^5+5N^4+4N^3-16N^2-8N+16}{2N(N+1)(N+2)^2},\label{JzAexpvalue2}\\
&\langle  J^z_{\text{B}} \rangle_{\text{ss}}(N)=\frac{N(N+1)(N^2-12)+8}{N(N+1)(N+2)^2}.\label{JzBexpvalue2}
\end{align} 
The negative-temperature state relaxation emerges from $N=4.$ In the limit $N\rightarrow\infty,$ the spin polarization in domain $D_{\text{A}}$ is $-N/2$
 whereas the spin polarization in domain $D_{\text{B}}$ becomes $1$.
Hence, the domain $D_{\text{A(B)}}$ is in the ground (excited) state.  
We will not calculate the logarithmic negativity \eqref{negativity} and just examine whether the quantum entanglement is generated or not between the two spin domains: the steady-state density matrix \eqref{steadystatedm4} consists of the separable-state part (the terms in the first, second, and third lines) plus the additional terms (from fourth to seventh lines).
Thus, the quantum entanglement is generated between the spin domains.

Now let us predict the formula for the density matrix at the steady state in the direct-sum spin space representation for general $N_{\text{B}}$.
By observing the steady-state density matrix structure \eqref{steadystatedm1} and \eqref{steadystatedm3} (note that the density matrix \eqref{steadystatedm1} and \eqref{steadystatedm3}
are for $N_{\text{B}}=1$ and 2, respectively), we see that in the direct-sum spin space the density matrix at the steady state has a structure such that only the end points in the diagonal blocks survive.
To explain this in a little more detail, let us denote the diagonal blocks for the density matrix as B$_1$, B$_2$, $\ldots$, and B$_{N_{\text{B}}+1}.$ 
Initially, in each block there is an element having finite value. Then by analyzing the equations of motion for the matrix elements,
at the steady state we may predict that only the end point in each block takes finite value and is equal to that of the element which was initially finite.  
This is because the subspaces $V_{j}$ ($j=(N_{\text{A}}+N_{\text{B}})/2,\ldots,(N_{\text{A}}-N_{\text{B}})/2$), which construct the direct-sum spin space, are orthogonal to each other.
Therefore, at the steady state the density matrix would have a structure 
\begin{align}
\rho_{\text{ss}}(N)=\sum_{i=1}^{N_{\text{B}}+1}P_{i}\Big{|} j_i;-j_i\rrangle[\Big]  \llangle[\Big] j_i;-j_i\Big{|}, \label{steadystatedm5}
 \end{align}
where $j_1=(N_{\text{A}}+N_{\text{B}})/2,j_2=(N_{\text{A}}+N_{\text{B}})/2-1,\ldots$ and $j_{N_{\text{B}}+1}=(N_{\text{A}}-N_{\text{B}})/2$.
The coefficients $P_{i}$ satisfy the conditions $0\leq P_{i}<1$ and $\sum_{i=1}^{N_{\text{B}}+1}P_{i}=1.$
 The matrix form of the steady state \eqref{steadystatedm5} is 
\begin{equation} 
	{\rho}_{\mathrm{ss}} (N) = 
	\left (
		\begin{array}{ccc|c|ccc} 
		 0 & & & & & & \\ 
		 & \ddots & & \cdots & \multicolumn{3}{c}{\mbox{\smash{\huge{$0$}}}} \\ 
		 & & P_{1} & & & & \\ \hline
		 \multicolumn{3}{c|}{\vdots} & \ddots & \multicolumn{3}{c}{\vdots} \\ \hline 
		 & & & & 0 & & \\
		 \multicolumn{3}{c|}{\mbox{\smash{\huge{$0$}}}} & \cdots & & \ddots & \\  
		 & & & & & & P_{N_{\mathrm{B}} +1} 
		\end{array}
	\right ).
\end{equation} 
The formula \eqref{steadystatedm5} is physically natural, because at zero temperature 
the total spin should relax so that the steady state must be described by the eigenstates whose eigenvalues take the minimum in the direct-sum spin subspaces which they belong to.
Indeed, the formula \eqref{steadystatedm5} satisfies the master equation \eqref{smasterequation1} as a steady-state solution.
By using the Clebsh-Gordan coefficients and describe the steady state \eqref{steadystatedm5} in the tensor-product subspace, we can discuss the two spin polarizations and whether the quantum entanglement is generated or not between 
the two domains. 

\section{Discussion and Conclusions}\label{discussionconclusions}
In this paper, we have investigated the dynamics of density matrix and its structure for the collective spin relaxation in the double spin-domain system.  In this system, the two spin domains couple equivalently to the common reservoir and the Hamiltonian is described by the total spin.  At initial time the spin ensemble in the first domain is in the excited state (all-up spin state) whereas the second spin ensemble to be in the ground state (all-down spin state) with the first spin size much larger than the second one.  The initial state does not have a full spin symmetry but is symmetric in each domain. 
Due to the angular-momentum conservation in the total system, the total system preserve the symmetry the initial state contains through the relaxation process. To analyze the spin relaxation process, the direct-sum spin space (direct sum of symmetric and asymmetric spaces) was more effective than the tensor-product representation.  This representation allowed us to reduce the dimensionality of the relevant Hilbert space significantly, and hence it became possible for the system to be analytically tractable.  For more complicated initial states, we may need to increase the dimensionality of the relevant Hilbert space, however, this methods will be also effective and beneficial for both analytical and numerical calculations.  

By analyzing the dynamics of the density-matrix elements in the direct-sum spin space, we have found that the density matrix for the collective spin relaxation had the following structure. 
The behavior of the density matrix shows that the populations in the symmetric space decades to its ground state, i.e. all spins are down, gradually losing the coherence between the symmetric and asymmetric subspaces.  The behavior in the asymmetric space is the same, although some excitations in spins remains in its ground states. When we see this behavior in the composite picture (the tensor-product space), the second domain which started at its ground state (the spin down) will be relaxed to the excited state.  The degree of the excitation is dependent on the difference of the spin domains in their size.  
For instance, in the case of second spin number equal to one, when the number of spins in the first domain is greater than two, 
the spin in the second domain decays to populate more than 50\% in the excited state, which indicates an effective negative temperature.
As the first spin number becomes sufficiently larger, the second spin domain is (almost) completely in the excited state where all the spins are up.  

The spin polarizations for both domains show the monotonic behaviors as functions of the first spin size in an opposite way.
The quantum entanglement between the two domains exhibit the non-monotonic behavior as a function of the first spin size. 
It is an increasing function when the first spin size is in the range from one to five. Then when it becomes equal to six and start to exceed, the quantum entanglement decreases monotonically and converges to zero.  
This behavior is consistent with the fact that when the first spin size is sufficiently large the steady state becomes separable with the first spin domain to be all down and the second spin domain to be all up.  
Correspondingly, the purity becomes one at first spin number to infinity.

These collective phenomena never occur in the single spin domain system and must be the consequence of the asymmetry of the spin state and the coupling to the common reservoir. 
The candidate hybrid quantum systems to observe these phenomena are, one is the GaAs semiconductor where nuclear spins are coupling to the electron spins in the QH state through the hyperfine interaction.
When we initially prepared the nuclear double spin domain having antiparallel configuration induced by the DNP \cite{DNP1,DNP2}, by tuning the QH state so that the linear dispersing NG mode as the bosonic reservoir emerge \cite{Kumadaetal,Fauziprb,yhamaetal}, 
we observe our collective phenomena.
The second candidate is the electron spin ensemble in the NV center in diamonds coupling to the superconducting resonator \cite{thomas}.

The interesting point of these two collective phenomena is that the characteristics of the steady state (the spin polarizations and the amount of quantum entanglement) are rather opposite to those at the initial time, although the steady states exhibit dependency to their initial states.
This relaxation behavior can be interesting to apply to quantum state manipulation and quantum information processes.  
Usually, the decoherence induced by the reservoir is regarded as an obstacle to perform the quantum information processing destroying the initial information of the system.  In these systems, after the system completely relaxed, the system has some in-print of the information initially the system has had.  This property may be exploited to implement robust quantum state manipulation.

\acknowledgements
This work was supported in part by the MEXT Grant-in-Aid for Scientic Research on Innovative Areas KAKENHI Grant Number JP15H05870 (Y.~H, E.~Y, and K.~N)
and MEXT Grant-in-Aid for Scientic Research(S) KAKENHI Grant Number JP25220601 (E.~Y and K.~N).

\appendix
\section{Mathematical Proof of Eq. \eqref{steadystatedm1}}\label{Nvs1steadystateproof}
In this section, we demonstrate the mathematical proof of Eq.  \eqref{steadystatedm1} by decomposing into three parts. Part I is the discussion for the dynamics of the diagonal elements whereas part II and III are those for the off-diagonal elements.

({\it Part I.  Diagonal elements}): \\
First, let us look at the dynamics of diagonal elements $\rho_{\alpha_{\text{I}},\alpha_{\text{I}}}$ in the block B$_1$ through the equation of motion \eqref{eom1}. 
For $\alpha_{\text{I}}=1$,  since the term $\rho_{\alpha_{\text{I}}-1,\alpha_{\text{I}}-1}$ does not exist the equation of motion \eqref{eom1} is described solely by $\rho_{1,1}$ as a linear differential equation.
Due to the initial condition \eqref{initialstate3}, we readily obtain $\rho_{1,1}$(t)$=0.$ Thus, Eq. \eqref{eom1} for $\alpha_{\text{I}}=2$
becomes the linear equation which is just described by $\rho_{2,2}$. From the initial condition \eqref{initialstate3}, we obtain 
\begin{align}
\left(\rho\right)_{2,2}(N,t)&=\frac{1}{N+1}\exp\Big{(} -4N\gamma t \Big{)}. \label{rho22solution}
 \end{align}
Next let us look at the Eq. \eqref{eom1} for $\alpha_{\text{I}}=3$, which is described by $\rho_{3,3}$ and $\rho_{2,2}$. 
Although we can solve this equation and obtained the solution $\left(\rho\right)_{3,3}(N,t)$ for any $t$, we just argue the steady-state solution because this is our interest.
At $t\rightarrow \infty$, both $\left(\dot{\rho}\right)_{3,3}(N,t)$ and $\left(\rho \right)_{2,2}(N,t)$ vanish. 
Thus, we have  $\left(\rho_{\text{ss}} \right)_{3,3}=0.$ By applying the same argument to other components sequentially, we have 
$\left(\rho_{\text{ss}} \right)_{4,4}=\ldots=\left(\rho_{\text{ss}} \right)_{N+1,N+1}=0.$ 
For $\alpha_1=N+2$, which is the end point of the block B$_1$, the right hand side of the equation is described solely by 
$\left(\rho \right)_{N+1,N+1}$ because the second term vanishes. Therefore, we have $\left(\rho_{\text{ss}} \right)_{N+2,N+2}=a_{\text{I}}=$const. 
This argument can be exactly applied for the dynamics of matrix elements $\rho^{\text{s}}_{\alpha_{\text{II}},\alpha_{\text{II}}}$ in the block B$_2$ using the equation of motion \eqref{eom2}. 
We obtain $\left(\rho_{\text{ss}} \right)_{2N+2,2N+2}=a_{\text{II}}=$const, which is finite and the rest of the components are zero. 

({\it Part II.  Off-Diagonal elements}-1):

We discuss the dynamics of off-diagonal elements $\rho_{\alpha_{\text{I}},\alpha_{\text{II}}}$ in the Block B$_3$ using the equation of motion \eqref{eom3}.  
The elements we consider are $\left(\rho\right)_{2,N+3}(N,t)$ and related ones. The matrix element $\left(\rho\right)_{2,N+3}(N,t)$ is the only off-diagonal element having a finite value at initial time.
We start from analyzing the dynamics of $\left(\rho\right)_{2,N+3}(N,t)$. 
Since $\left(\rho \right)_{1,N+2}(N,t)$ belongs to the block B$_1$, the term $\rho_{\alpha_{\text{I}} -1,\alpha^\prime_{\text{II}} -1}$ in Eq. \eqref{eom3} vanishes.
Thus, the equation of motion \eqref{eom3} for $\alpha_{\text{I}} =2,\alpha^\prime_{\text{II}}=N+3 $ is solely described by $\left(\rho \right)_{2,N+3}(N,t)$.
From the initial condition \eqref{initialstate3}, it is solved as
\begin{align}
\left(\rho \right)_{2,N+3}(N,t)&=-\frac{\sqrt{N}}{N+1}\exp\Big{(} -(3N-1)\gamma t \Big{)}. \label{rho2N+3solution}
 \end{align}
Next, what we do is we repeat exactly the same argument which we did in Part I.
Here again, we just consider only the steady-state solutions. For $\alpha_{\text{I}} =3,\alpha^\prime_{\text{II}}=N+4$ the right-hand side of equation of motion \eqref{eom3} is described by $\left(\rho^{\text{s}}\right)_{3,N+4}(N,t)$ and $\left(\rho \right)_{2,N+3}(N,t)$.
 From Eq. \eqref{rho2N+3solution}, we see that the steady-state solution for $\left(\rho \right)_{2,N+3}(N,t)$ is zero. Therefore, the steady-state solution for $\left(\rho \right)_{3,N+4}(N,t)$ is also zero.
We repeat this argument sequentially for  $\alpha_{\text{I}} =4,\alpha^\prime_{\text{II}}=N+5,\ldots,\alpha_{\text{I}} =N+1,\alpha^\prime_{\text{II}}=2N+2$. Then we have 
$\left(\rho_{\text{ss}} \right)_{4,N+5}=\ldots=\left(\rho_{\text{ss}} \right)_{N+1,2N+2}=0.$ As a result, the off-diagonal components for $\alpha_{\text{I}} =3,4,\ldots,N+1,\alpha^\prime_{\text{II}}=N+4,N+5,\ldots,2N+2$
vanish at the steady state. Such behaviors are consistent with the plots in Fig. \ref{DMoffdiagonal}.

({\it Part III.  Off-Diagonal elements}-2):  \\
In this part, we discuss the dynamics of off-diagonal elements ${\rho}_{\alpha_{\text{I}},\alpha^\prime_{\text{I}}}$ and ${\rho} _{\alpha_{\text{II}},\alpha^\prime_{\text{II}}}$, and ${\rho} _{\alpha_{\text{I}},\alpha_{\text{II}}}$
which were not discussed in the Part II. Since the arguments for ${\rho} _{\alpha_{\text{I}},\alpha^\prime_{\text{I}}}$, ${\rho} _{\alpha_{\text{II}},\alpha^\prime_{\text{II}}}$, and ${\rho} _{\alpha_{\text{I}},\alpha_{\text{II}}} $ become exactly the same, here we will just present the argument for ${\rho} _{\alpha_{\text{I}},\alpha^\prime_{\text{I}}}$. These elements are the simplest cases to analyze the steady-state solution because from  Eq. \eqref{initialstate3} all these components are zero at the initial state.
First, we start with the dynamics of ${\rho} _{1,\alpha^\prime_{\text{I}}}$ ($\alpha^\prime_{\text{I}}>1$). 
From the equation of motion \eqref{eom1} and the initial condition \eqref{initialstate3}, we have $({\rho})_{1,\alpha^\prime_{\text{I}}}(N,t)=0$.
As we mentioned above, since all the components at initial time are zero, we can easily show that $(\rho)_{2,\alpha^\prime_{\text{I}}+1}(N,t)=({\rho})_{3,\alpha^\prime_{\text{I}}+2}(N,t)\ldots=({\rho})_{N+3-\alpha^\prime_{\text{I}},N+2}(N,t)=0.$
Similarly, from the equations of motion \eqref{eom2}, \eqref{eom3}, and the initial condition \eqref{initialstate3}, all the matrix elements ${\rho}_{\alpha_{\text{II}},\alpha^\prime_{\text{II}}}$ and ${\rho}_{\alpha_{\text{I}},\alpha_{\text{II}}} $ under the consideration are zero.
Therefore, all these off-diagonal elements vanish at the steady state. This is consistent with the results shown in Fig. \ref{DMoffdiagonal}.

As a result, all the off-diagonal elements vanish at the steady state. The only finite elements are $\rho_{N+2,N+2}$ and $\rho_{2N+2,2N+2}$.
With taking account of the constraint Tr($\rho_{\text{ss}}(N)$)=1, the natural choices for  $\left( \rho_{\text{ss}}(N)\right)_{N+2,N+2}=a_{\text{I}}$  
and $\left( \rho_{\text{ss}}(N)\right)_{2N+2,2N+2}=a_{\text{II}}$ are 
\begin{align}
a_{\text{I}}=\frac{1}{N+1}, \quad 
 a_{\text{II}}=\frac{N}{N+1}. \label{sssolution}
\end{align}  
This is because the symmetric subspace and asymmetric subspace are orthogonal to each other. There must be no spin population transfer between them. 
In other words, the probability weight for each spin subspace must be invariant under the time evolution. Indeed, this is what we see in Fig. \ref{DMdiagonal}.
Consequently, we obtain the steady-state formula \eqref{steadystatedm1}.

\end{document}